\documentclass[10pt,a4paper]{article}

\usepackage{arxiv}
\usepackage[utf8]{inputenc}
\usepackage[T1]{fontenc}
\usepackage{amsmath,amssymb,amsfonts}
\usepackage{graphicx}
\usepackage{booktabs}
\usepackage{xcolor}
\usepackage{url}
\usepackage{tikz}
\usetikzlibrary{arrows.meta,shapes.geometric,shapes.multipart,positioning,fit,backgrounds,calc}
\usepackage{float}
\usepackage{placeins}
\usepackage[numbers,sort&compress]{natbib}
\usepackage{pdfpages}
\usepackage{hyperref}
\usepackage[nameinlink,capitalize]{cleveref}

\graphicspath{{figures/}}

\title{Differentiable Inverse Design of Short-Range Order in High-Entropy Alloys:\\
From Target $\alpha$ to Target Property}

\author{%
  Tiancheng Ding\textsuperscript{1} \quad
  Conrard Giresse Tetsassi Feugmo\textsuperscript{1,2,*}\\[4pt]
  \small\textsuperscript{1}Department of Chemistry, University of Waterloo,
  Waterloo, Ontario N2L 3G1, Canada\\
  \small\textsuperscript{2}Department of Physics and Astronomy, University of Waterloo,
  Waterloo, Ontario N2L 3G1, Canada\\[2pt]
  \small\textsuperscript{*}Corresponding author: \href{mailto:cgtetsas@uwaterloo.ca}{cgtetsas@uwaterloo.ca}
}

\date{}

\begin{document}

\maketitle

\begin{abstract}
Short-range order (SRO) governs the mechanical response of multi-principal-element
alloys, yet inverse design from a target Warren--Cowley signature to an atomic
configuration, and onward to a target property, is typically solved in disconnected
stages that cannot share gradients. We replace stochastic atomistic reverse
Monte Carlo (ARMC) with end-to-end gradient descent on a continuous-occupancy
lattice and connect it to a differentiable property surrogate trained on
real interatomic-potential elastic constants. Three contributions are reported:
(1)~gradient-based SRO-to-structure inversion matches ARMC accuracy at small
cell sizes, is $6\times$ faster and $8\times$ more accurate on 4000-atom
supercells, and scales smoothly to $K$ species with no per-move bookkeeping;
(2)~a thermodynamic regulariser derived from the Non-Interacting Molecule
Method (NIMM) of Rao and Curtin, generalised here from binary to $K$ species,
anchors designed configurations to the free-energy surface of a chosen
effective pair interaction, removing statistically valid but physically
inconsistent solutions;
(3)~a descriptor multilayer perceptron trained on composition, multi-shell
$\alpha$, and occupancy statistics drives closed-loop property design: applied
to a nine-system survey of face-centred cubic (FCC) and body-centred cubic (BCC)
alloys, the pipeline recovers
SRO-induced $C_{11}$ changes from $-20\,\%$ (Cr--Cu--Ni) to $+57\,\%$
(Cr--Fe--Ni) in the 108-atom survey cells, with cell-size verification
($3\times3\times3$ to $6\times6\times6$) showing that the sign and
magnitude of $\Delta C_{11}$ are not converged below 864 atoms, and
closes the design loop on real LAMMPS modified embedded atom
method (MEAM) elastic constants for Co--Cr--Ni with three of four target
$C_{11}$ values within $6\,\%$. The framework is released as the
open-source PyTorch package \texttt{anisro} and provides a path toward
gradient-based property design in concentrated solid solutions.
\end{abstract}

\textbf{Keywords:} high-entropy alloys; short-range order; inverse design;
differentiable programming; elastic constants

\section{Introduction}
\label{sec:intro}

Chemical short-range order (SRO) is emerging as a controllable design
variable in concentrated solid solutions and high-entropy alloys
(HEAs) \citep{george_2019_natrevmater,cantor_2004_msea,yeh_2004_aem,miracle_2017_acta}.
The Warren--Cowley parameter $\alpha_{ij}^{(r)}$, a signed measure of
hetero-pair enrichment in the $r$-th coordination shell relative to a
random alloy, sets the local bonding topology and, through it,
the generalized stacking-fault energy surface, dislocation core width,
twinning stress, and hydrogen trapping landscape. In the canonical
Cr--Co--Ni medium-entropy alloy, atom-probe tomography and
electron-diffraction experiments revealed nanoscale ordered domains
associated with a $15$--$20\,\%$ shift in stacking-fault energy and a
transition in deformation mechanism from dislocation glide to nanotwinning
\citep{zhang2020nature,ding2018pnas,zhou_2022_atomic_sro_crconi}. Subsequent studies across
face-centred and body-centred chemistries have tied SRO to yield
strength, work-hardening capacity, hydrogen embrittlement resistance,
and irradiation tolerance \citep{cao2021npj,chen2021nature,wu2021science,li_2019_natcomm,wu_2021_sro_mechanical,chen_2022_structure_motif}.
For alloy designers, the implication is concrete: $\alpha_{ij}^{(r)}$ is
no longer merely a thermodynamic descriptor but a target variable whose
value, if achievable in the laboratory or in simulation, predicts the
mechanical signature of the alloy.

The microscopic mechanism connecting SRO to elastic response runs through
the local force-constant environment.
A negative $\alpha_{ij}^{(r)}$ signals hetero-pair enrichment in shell $r$:
the coordination shell of species $i$ contains more $j$ neighbours than
a random alloy of the same composition would.
When the $i$--$j$ bond is stiffer than the $i$--$i$ or $j$--$j$ alternatives,
as in Cr--Fe pairs within the Cr--Fe--Ni system, this enrichment
raises the effective interatomic force constants and stiffens the crystal,
manifesting as an increase in $C_{11}$.
Conversely, when the hetero-pair is softer or introduces lattice strain
through size mismatch, as in Co--Cr pairs within Co--Cr--Ni, ordering
softens the crystal and reduces $C_{11}$.
The quantitative magnitude depends on the species contrast in bond stiffness,
the shell coordination geometry, and the degree of ordering: even modest
$|\alpha| \sim 0.3$--$0.4$ values, well within the pair-cluster validity
band, can shift elastic moduli by tens of GPa.

The computational tools needed to quantify and set SRO have matured
rapidly. On the forward side, density-functional theory combined with
canonical Monte Carlo remains the reference for small supercells and
short trajectories \citep{tamm2015actamat,fernandez2020prm,metropolis_1953}; machine-learning
interatomic potentials (MLIPs) trained on DFT data extend accessible
system sizes to hundreds of thousands of atoms \citep{ko2021npj,deng2023npj,batzner_2022_nequip,musaelian_2023_allegro,han_2025_mlip_sro,liu_2019_ml_sro};
and the analytical Non-Interacting Molecule Method (NIMM) of Rao and Curtin
\citep{rao2022actamat} delivers closed-form Warren--Cowley parameters from
a small set of effective pair interactions (EPIs) via an inexpensive Newton
solve, with a delineated validity band. On the inverse side, generating
an atomic configuration whose statistical $\alpha$ matches a prescribed
target tuple, atomistic reverse Monte Carlo (ARMC) is the de facto
standard, with the recent work of Sheriff and co-workers establishing best
practice for multi-component systems \citep{sheriff2024pnas,sheriff_2024_motif_gnn,jiang2016prb}.

Despite these individual advances, the chain from a target $\alpha$ to a
target mechanical property remains stitched together by discrete stages that
cannot exchange gradients. ARMC produces configurations consistent with a
target $\alpha$ but provides no smooth gradient with respect to that target
or to the underlying EPI; coupling its output to a property predictor and
propagating sensitivities backward requires either a derivative-free outer
loop or piecewise re-training. Meanwhile, the broader community of
differentiable molecular simulation has demonstrated what is possible when
forward and inverse operators are written as composable, automatically
differentiable programs: end-to-end energy minimisation, NVT sampling, and
force-field optimisation have all been demonstrated in JAX or PyTorch
\citep{schoenholz2020jaxmd,doerr2021npj,batatia2022mace,paszke_2019_pytorch,goodrich_2021_diff_selfassembly,allen_2022_diff_learned_sim}. To our knowledge,
no equivalent end-to-end-differentiable formulation exists for the
SRO-to-structure-to-property problem in HEAs.

We close this gap. The core technical move is to write the Warren--Cowley
counter itself as a pure-functional, fully differentiable tensor operation
acting on a continuous per-site species probability $P \in [0,1]^{N \times K}$
parameterised by free logits $\boldsymbol{\ell}$ via a softmax. With this
representation, gradient descent on $\boldsymbol{\ell}$ minimises the same
$\sum(\hat\alpha - \alpha^\star)^2$ residual that ARMC minimises by stochastic
swap moves, while simultaneously respecting composition, binary-entropy, and
thermodynamic regularisers. The thermodynamic regulariser is a differentiable
reformulation of the NIMM free energy generalised to $K$ species; it anchors
the designed configuration to the free-energy surface of a chosen EPI and
removes statistically valid but physically inconsistent solutions. The same
logit parameterisation extends to property design: replacing the SRO loss
with a surrogate property prediction connects the target mechanical property
to the atomic configuration via a single backward pass.

Three contributions are reported. Gradient SRO-to-structure inversion matches
ARMC accuracy on small binary cells and is $6\times$ faster and $8\times$
more accurate on 4000-atom supercells, with smooth scaling to $K$ species
because the per-step cost is independent of species count. A NIMM-derived
free-energy regulariser, generalised here from binary to $K$ species
following Rao and Curtin~\citep{rao_curtin_2022_nimm}, makes designed
configurations physically consistent with a chosen EPI surface rather than
merely statistically $\alpha$-matched. A descriptor multilayer perceptron
trained on real LAMMPS-MEAM elastic constants drives closed-loop property
design on Co--Cr--Ni and recovers SRO-induced $C_{11}$ changes ranging from
$-20\,\%$ (Cr--Cu--Ni) to $+57\,\%$ (Cr--Fe--Ni) across a nine-system
FCC/BCC survey. The complete pipeline is illustrated in
Figure~\ref{fig:pipeline}. The framework is released as the open-source
PyTorch package \texttt{anisro}.

\section{Methods}
\label{sec:methods}

\begin{figure}[H]
  \centering
  \includegraphics[width=0.97\textwidth]{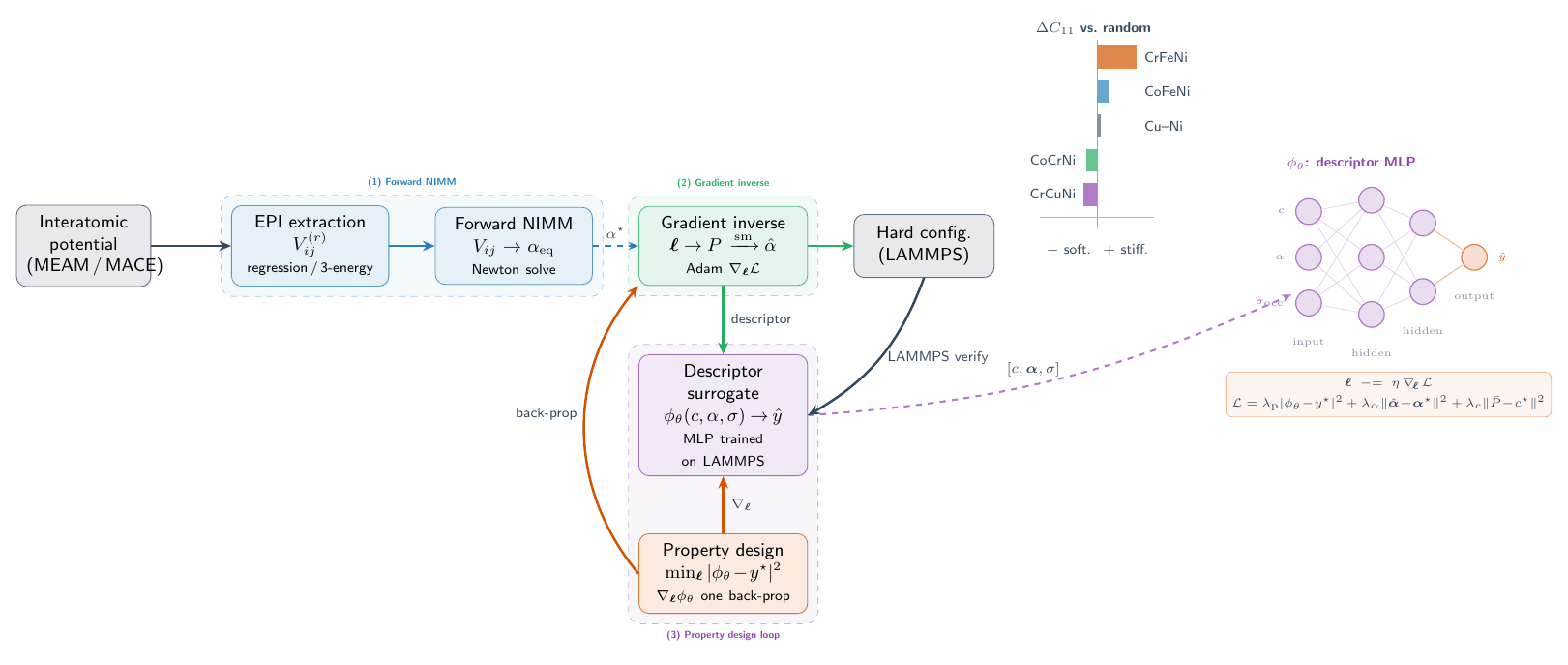}
  \caption{ANISRO overview. ANISRO connects any interatomic potential to
  gradient-based design of atomic short-range order and target mechanical
  properties. Stage~1 (blue): EPI extraction and forward NIMM prediction of
  equilibrium $\alpha$. Stage~2 (green): differentiable SRO-to-structure
  inversion via soft site occupancies. Stage~3 (purple/orange): descriptor
  surrogate and closed-loop property design with gradients flowing end-to-end.
  The mini bar chart shows the range of SRO-induced $\Delta C_{11}$ across
  the nine surveyed alloy systems.\label{fig:anisro-overview}}
\end{figure}

Figure~\ref{fig:anisro-overview} provides a high-level overview of the ANISRO framework; the complete pipeline, with all four stages colour-coded, is assembled in Figure~\ref{fig:pipeline} at the close of this section.

\subsection{Warren--Cowley short-range order parameters}
\label{ssec:wc-definition}

We adopt the Warren--Cowley convention throughout \citep{warren_1950,cowley_1950}. For an $N$-site lattice
with $K$ species and the $r$-th coordination shell of coordination number
$z_r$, the joint pair probability $P_{ij}^{(r)}$ is the probability that
a site of species $i$ has a species-$j$ neighbour in shell $r$. The
Warren--Cowley parameter is
\begin{equation}
  \alpha_{ij}^{(r)} = 1 - \frac{P_{ij}^{(r)}}{c_j},
  \label{eq:wc-def}
\end{equation}
where $c_j$ is the global mole fraction of species $j$. A value of
$\alpha_{ij}^{(r)} < 0$ denotes enrichment of $j$ around $i$ (ordering tendency);
$\alpha_{ij}^{(r)} > 0$ denotes depletion (clustering or phase-separation
tendency). For a random alloy $\alpha_{ij}^{(r)} = 0$ for all pairs and shells.
The matrix $\boldsymbol{\alpha}^{(r)} \in \mathbb{R}^{K\times K}$ is symmetric
($\alpha_{ij}^{(r)} = \alpha_{ji}^{(r)}$) and satisfies the linear constraint
$\sum_j c_j \alpha_{ij}^{(r)} = 0$ for every $i$, leaving $K(K-1)/2$ independent
hetero-pair parameters per shell.

\subsection{Forward NIMM solver for $K$ species}
\label{ssec:forward-nimm}

The forward operator maps a per-shell hetero-pair effective pair interaction
(EPI) table $\{V_{ij}^{(r)}\}$, the composition $\{c_i\}$, and the temperature
$T$ to the Warren--Cowley matrix $\{\alpha_{ij}^{(r)}\}$. The EPI is defined in
the formation-energy convention:
\begin{equation}
  V_{ij}^{(r)} = U_{ii}^{(r)} + U_{jj}^{(r)} - 2 U_{ij}^{(r)},
  \label{eq:epi-def}
\end{equation}
where $U_{ij}^{(r)}$ is the bond energy per site for a fully ordered
$ij$-nearest-neighbour environment on shell $r$. A positive $V_{ij}^{(r)}$
lowers the energy of $ij$ bonds relative to the symmetric mean, driving
ordering ($\alpha_{ij} < 0$); a negative $V_{ij}^{(r)}$ promotes clustering
($\alpha_{ij} > 0$).

We start from the Non-Interacting Molecule Method of Rao and Curtin
\citep{rao2022actamat} and generalise their binary derivation to $K$ species.
For each coordination shell $r$, the configurational state of a pair cluster
is one of $K$ homo categories $(i,i)$ with degeneracy $g_k = 1$ and
$K(K-1)/2$ hetero categories $(i,j)$, $i < j$, with $g_k = 2$, giving
$K_r = K(K+1)/2$ cluster classes in total. Defining the dimensionless pair
fugacity
\begin{equation}
  x_{ij}^{(r)} = \exp\!\left(\frac{-V_{ij}^{(r)}}{k_B T}\right),
  \qquad x_{ii}^{(r)} = 1,
  \label{eq:fugacity}
\end{equation}
the variational stationarity condition from maximising the cluster-occupancy
entropy subject to mass balance gives the per-cluster log-probability:
\begin{equation}
\ln p_k^{(r)} = \ln g_k
        - \frac{1}{2}\sum_{i,j} N_{ij,k}^{(r)}\ln x_{ij}^{(r)}
        + \sum_i \lambda_i\!\left(2 N_{ii,k}^{(r)} + \sum_{j\neq i} N_{ij,k}^{(r)}\right)
        + \mu,
\label{eq:stationarity}
\end{equation}
where $N_{ij,k}^{(r)} \in \{0,1\}$ is the multiplicity of pair $(i,j)$ in
cluster class $k$, the Lagrange multipliers $\{\lambda_i\}$ enforce the
$K$ mass-balance constraints
\begin{equation}
  2\sum_k p_k N_{ii,k}^{(r)} + \sum_{j\neq i}\sum_k p_k N_{ij,k}^{(r)} = 2c_i
  \quad \forall\, i,
  \label{eq:mass-balance}
\end{equation}
and $\mu$ enforces $\sum_k p_k = 1$. The sign of the pair fugacity term
$-\tfrac{1}{2}\sum_{ij} N_{ij,k} \ln x_{ij}$ combines with the $-1$ from
the Stirling chain rule to yield $-\ln x_{ij}/2$ in the final stationarity
equation, so that a positive $V_{ij}$ (positive $\ln x_{ij}$) increases
$\ln p_k$ for the heteropair class and drives $\alpha_{ij} < 0$. This
sign convention is numerically validated in Section~\ref{ssec:res-nimm}
and formally derived in Section~S1 of the Supplementary Information.

The $K_r + K + 1$ nonlinear equations are solved shell-by-shell by a
damped Newton iteration with column-rescaled Jacobian:
\begin{equation}
  \mathbf{J}\,\delta\mathbf{x} = -\mathbf{r},\qquad
  \mathbf{J}_{kl} = \frac{\partial r_k}{\partial x_l},
  \label{eq:newton}
\end{equation}
where the residual $\mathbf{r}$ collects the stationarity equations
\eqref{eq:stationarity}, the mass-balance constraints \eqref{eq:mass-balance},
and the normalisation constraint. An Armijo line search keeps every $p_k > 0$
throughout the iteration. Newton residuals reach machine precision
($\|\mathbf{r}\| < 10^{-14}$) in fewer than 30 iterations for all systems
in this work, including the five-species Cantor alloy ($K_r = 15$).
The full algorithm with pseudocode and convergence data is given in
Section~S1.2 of the Supplementary Information.

The joint pair probability and Warren--Cowley parameter follow from
\begin{align}
  P_{ij}^{(r)} &= \sum_k p_k^{(r)}\, N_{ij,k}^{(r)} / N_r, \label{eq:joint-pair}\\
  \alpha_{ij}^{(r)} &= 1 - P_{ij}^{(r)} / (2 c_i c_j), \label{eq:wc-from-pij}
\end{align}
matching Eqs.~(12) and (13) of Rao and Curtin.
Here $P_{ij}^{(r)}$ counts symmetric pairs (both $(i,j)$ and $(j,i)$ directions),
so $P_{ij}^{(r)} = 2c_ic_j$ for a random alloy, which recovers $\alpha_{ij}^{(r)}=0$
consistently with Eq.~\eqref{eq:wc-def}; this differs from the single-direction
conditional probability used in Eq.~\eqref{eq:wc-def} by the factor $2c_i$.
The solver is implemented in
PyTorch (\texttt{anisro.forward.pair\_cluster}) so its outputs remain
differentiable through the EPI table and composition, enabling gradient-based
EPI regression (Section~\ref{ssec:epi-extraction}).

The NIMM model is quantitatively reliable in the pair-cluster validity band
$0.5 < x_{ij}^{(r)} < 1.5$, equivalently $|V_{ij}^{(r)}| \lesssim 0.7\,k_BT$,
as delineated by Rao and Curtin. Several of the dedicated 2NN-MEAMs studied here
produce EPIs at 300\,K that violate this bound; in those cases the gradient
inverse targets Metropolis MC $\alpha$ directly, using NIMM only as a soft prior
rather than as a hard constraint.

\subsection{Differentiable Warren--Cowley counter}
\label{ssec:differentiable-alpha}

\begin{figure}[H]
  \centering
  \includegraphics[width=0.88\textwidth]{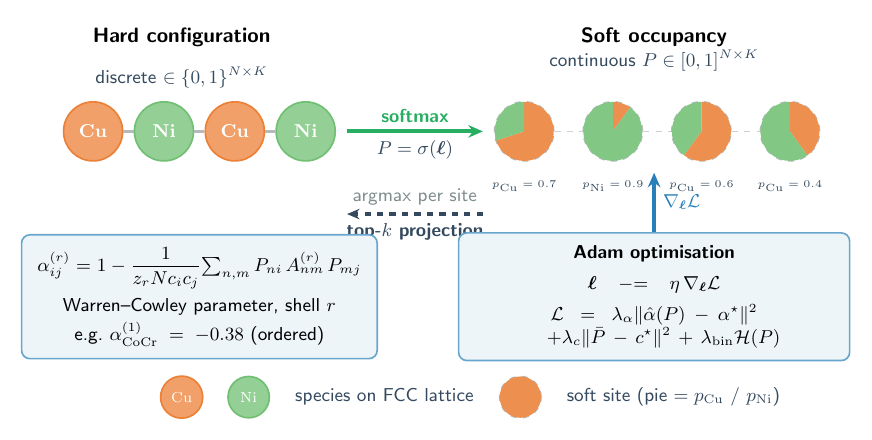}
  \caption{Concept of the differentiable SRO-to-structure inverse.
  Each lattice site carries a logit vector $\boldsymbol{\ell}_i \in \mathbb{R}^K$;
  passing through a per-site softmax yields soft occupancy probabilities
  $P_i = \sigma(\boldsymbol{\ell}_i)$. The differentiable $\alpha$ counter
  evaluates Warren--Cowley parameters from $P$, enabling gradient descent
  on the SRO loss. The hard top-$k$ projection converts the soft optimum to
  a discrete LAMMPS configuration.\label{fig:inverse-schematic}}
\end{figure}

For the inverse problem, atomic configurations are parameterised by a soft
occupancy tensor $P \in [0,1]^{N \times K}$ with $\sum_i P_{ni} = 1$ for every
site $n$ (distinct from the joint pair probability $P_{ij}^{(r)}$ of
Section~\ref{ssec:wc-definition}; context distinguishes the two), obtained from a free logit matrix $\boldsymbol{\ell} \in \mathbb{R}^{N \times K}$
via row-wise softmax:
\begin{equation}
  P_{ni} = \frac{\exp(\ell_{ni})}{\sum_{i'}\exp(\ell_{ni'})}.
  \label{eq:softmax}
\end{equation}
When $\boldsymbol{\ell}$ is highly peaked, $P$ approaches a one-hot matrix
(integer occupancy); gradient descent on $\boldsymbol{\ell}$ explores a
continuous relaxation of the discrete combinatorial space.
Figure~\ref{fig:inverse-schematic} illustrates this parameterisation and the hard projection step.

Given shell adjacency matrices $A^{(r)} \in \{0,1\}^{N \times N}$ with
$\sum_m A^{(r)}_{nm} = z_r$ for every site $n$, the number of $(i,j)$ pairs
in shell $r$ is
\begin{equation}
  \mathrm{num}_{ij}^{(r)} = \sum_{n,m} P_{ni}\, A^{(r)}_{nm}\, P_{mj}
                          \equiv \texttt{einsum(`ni,rnm,mj->rij', P, A, P)},
  \label{eq:diff-counter-num}
\end{equation}
the soft composition of species $i$ is $n_i = \sum_n P_{ni}$, $c_i = n_i/N$,
and the differentiable Warren--Cowley parameter is
\begin{equation}
  \hat{\alpha}_{ij}^{(r)}(P) = 1 - \frac{\mathrm{num}_{ij}^{(r)}}{z_r\, n_i\, c_j}.
  \label{eq:diff-counter-alpha}
\end{equation}
This expression is exact when $P$ is one-hot, reducing to the discrete
statistical counter used by ARMC, and smooth for arbitrary $P \in [0,1]^{N\times K}$.
The einsum in Eq.~\eqref{eq:diff-counter-num} carries an optional leading
batch dimension unchanged, making the counter compatible with
\texttt{torch.func.vmap} for parallelism over multiple target-$\alpha$ tuples.

\subsection{Gradient SRO-to-structure inverse}
\label{ssec:gradient-inverse}

The inverse objective minimises the deviation of the soft configuration's
SRO from a target matrix $\boldsymbol{\alpha}^\star$. The SRO loss is a
shell-weighted masked mean-squared error over the $\Omega$ observed hetero-pairs:
\begin{equation}
  \mathcal{L}_{\alpha}(\boldsymbol{\ell}) =
  \sum_{r}\, w_r \sum_{(i,j)\in\Omega}
  \bigl(\hat{\alpha}_{ij}^{(r)}(P(\boldsymbol{\ell})) - \alpha_{ij}^{\star\,(r)}\bigr)^{2}.
  \label{eq:loss-alpha}
\end{equation}
Two regularisers absorb the hard constraints that ARMC enforces implicitly
through its move design:
\begin{align}
  \mathcal{L}_{\rm comp}(\boldsymbol{\ell})
    &= \sum_{i=1}^{K}\!\left(\overline{P}_i - c_i^{\star}\right)^{2},\quad
     \overline{P}_i = \frac{1}{N}\sum_n P_{ni},
  \label{eq:loss-comp}\\
  \mathcal{L}_{\rm bin}(\boldsymbol{\ell})
    &= -\frac{1}{NK}\sum_{n,i} P_{ni}\ln P_{ni}.
  \label{eq:loss-bin}
\end{align}
The composition loss $\mathcal{L}_{\rm comp}$ drives the mean occupation of each
species toward the target mole fraction; the binarisation loss
$\mathcal{L}_{\rm bin}$ is the per-site entropy, which is maximised by uniform
$P_{ni} = 1/K$ and minimised (to zero) by one-hot $P$; minimising
$-\mathcal{L}_{\rm bin}$ therefore pulls every site toward a definite species
assignment. The total loss is
\begin{equation}
  \mathcal{L}_{\rm total}
    = \lambda_\alpha\,\mathcal{L}_\alpha
    + \lambda_{\rm comp}\,\mathcal{L}_{\rm comp}
    + \lambda_{\rm bin}\,\mathcal{L}_{\rm bin}
    + \lambda_{\rm phys}\,\mathcal{L}_{\rm phys},
  \label{eq:loss-total}
\end{equation}
where $\mathcal{L}_{\rm phys}$ is the optional NIMM free-energy regulariser
(Section~\ref{ssec:nimm-regulariser}). Default weights used throughout:
$\lambda_\alpha = 1$, $\lambda_{\rm comp} = 5$, $\lambda_{\rm bin} = 0.01$,
$\lambda_{\rm phys} = 0$ (gradient inverse) or $0.1$ (hybrid).

Optimisation uses Adam \citep{kingma2015adam} in PyTorch \citep{paszke_2019_pytorch} with learning rate $10^{-2}$,
$(\beta_1,\beta_2) = (0.9, 0.999)$, and logit clipping to $[-10, 10]$.
After $N_{\rm steps} = 1500$ steps the best-loss snapshot is projected to a
hard configuration by per-species top-$k$ selection: for each species $i$,
the $\lfloor c_i^\star N \rfloor$ sites with highest $P_{ni}$ are assigned
species $i$. This projection introduces a soft-to-hard drift in $\alpha$;
we report both the soft $\hat{\alpha}$ at the end of optimisation and the
hard $\alpha$ computed on the projected configuration by the discrete counter.
The complete per-species top-$k$ algorithm is detailed in Section~S3 of
the Supplementary Information.

\subsection{NIMM free-energy regulariser}
\label{ssec:nimm-regulariser}

To bias the gradient inverse toward configurations that are thermodynamically
plausible under a chosen EPI, rather than merely statistically matching
a target $\alpha$, we add a differentiable approximation to the NIMM
Helmholtz free energy per atom. Generalising Eq.~(22) of Rao and Curtin to
$K$ species and fixed composition gives, per coordination shell $r$,
\begin{equation}
  \frac{F^{(r)}}{N} \;=\;
  -\frac{z_r}{4}\sum_{i \neq j} \tilde{P}_{ij}^{(r)}\, V_{ij}^{(r)}
  \;+\;
  \frac{k_B T\, z_r}{4}\sum_{i,j} \tilde{P}_{ij}^{(r)}\ln \tilde{P}_{ij}^{(r)},
  \label{eq:nimm-free-energy}
\end{equation}
where $\tilde{P}_{ij}^{(r)} = \mathrm{num}_{ij}^{(r)} / (z_r n_i)$ is the
joint pair probability obtained by normalising Eq.~\eqref{eq:diff-counter-num}
by the expected number of $i$-type bonds. The first term is the mean pair
energy (the heteropair contribution only; homo-pair energies drop out of the
variational problem at fixed composition); the second is the configurational
entropy, $-TS$. The minus sign on the pair-energy term follows the EPI sign
convention \eqref{eq:epi-def}: for $V_{ij} > 0$, minimising $F^{(r)}$ increases
$\tilde{P}_{ij}^{(r)}$ (more heteropairs), which is the same equilibrium
the forward Newton solver predicts via Eq.~\eqref{eq:stationarity}.

The physics regulariser is
$\mathcal{L}_{\rm phys} = \sum_r F^{(r)}/N$, and its gradient with respect to
$\boldsymbol{\ell}$ is computed by PyTorch autograd through
Eqs.~\eqref{eq:softmax}--\eqref{eq:nimm-free-energy}. When $\lambda_{\rm phys} > 0$
the optimiser minimises the combined SRO + thermodynamic loss, landing on
configurations that simultaneously match $\alpha^\star$ and sit near the
free-energy minimum of the chosen potential.

\subsection{Property descriptor and surrogate}
\label{ssec:surrogate}

The descriptor vector $\boldsymbol{d}(P)$ concatenates four groups of features
computed directly from the soft occupancy tensor:
\begin{equation}
  \boldsymbol{d}(P) = \Bigl[
    \underbrace{c_1,\ldots,c_K}_{\rm composition},\;
    \underbrace{\hat{\alpha}_{ij}^{(1)}}_{\rm 1NN\;SRO},\;
    \underbrace{\hat{\alpha}_{ij}^{(2)}}_{\rm 2NN\;SRO},\;
    \underbrace{\sigma^2_{\rm occ,\,i}}_{\rm occupancy\;var.}
  \Bigr] \in \mathbb{R}^{d},
  \label{eq:descriptor}
\end{equation}
where $\sigma^2_{{\rm occ},i} = \tfrac{1}{N}\sum_n (P_{ni} - c_i)^2$ measures
the per-species occupancy variance (large for well-segregated SRO, small for
diffuse). For a $K$-species, 2-shell system, $d = K + K(K-1) + K = K(K+1) = 12$
for $K = 3$.

\begin{figure}[H]
\centering
\resizebox{\linewidth}{!}{%
\begin{tikzpicture}[
  >=Stealth,
  node distance = 0.45cm and 1.0cm,
  every node/.style = {font=\small, align=center},
  box/.style        = {draw, rounded corners=4pt, inner sep=7pt, align=center},
  databox/.style    = {box, fill=blue!10,   draw=blue!50!black,   minimum width=2.2cm},
  compbox/.style    = {box, fill=green!10,  draw=green!50!black,  minimum width=2.6cm},
  mlpbox/.style     = {box, fill=orange!12, draw=orange!60!black, minimum width=2.4cm},
  trainbox/.style   = {box, fill=gray!12,   draw=gray!55,         minimum width=2.4cm},
  outbox/.style     = {box, fill=red!8,     draw=red!50!black,    minimum width=1.8cm},
  arr/.style        = {->, thick},
  darr/.style       = {->, thick, dashed, color=gray!60!black},
]

\node[databox] (logit)
  {Logit field \\ $\boldsymbol{\ell}\in\mathbb{R}^{N\times K}$};

\node[databox, right=of logit] (soft)
  {Soft occupancy \\ $P=\sigma(\boldsymbol{\ell})$};

\node[compbox, right=of soft] (desc)
  {Descriptor $\boldsymbol{d}(P)\in\mathbb{R}^{12}$ \\[3pt]
   composition $c_i$ \\
   1NN SRO $\hat{\alpha}_{ij}^{(1)}$ \\
   2NN SRO $\hat{\alpha}_{ij}^{(2)}$ \\
   occ.\ var.\ $\sigma^2_{{\rm occ},i}$};

\node[mlpbox, right=of desc] (mlp)
  {MLP $\phi_\theta$ \\ $12{\to}64{\to}64{\to}1$ \\ LayerNorm + ReLU};

\node[outbox, right=of mlp] (pred)
  {$\hat{y}$ \\ (e.g.\ $C_{11}$)};

\draw[arr] (logit) -- node[above,font=\scriptsize] {softmax} (soft);
\draw[arr] (soft)  -- node[above,font=\scriptsize] {Eq.~\eqref{eq:descriptor}} (desc);
\draw[arr] (desc)  -- (mlp);
\draw[arr] (mlp)   -- (pred);

\node[trainbox, below=1.4cm of desc] (lammps)
  {LAMMPS labels $y$ \\ (gradient-designed configs)};

\node[trainbox, right=of lammps] (loss)
  {MSE loss \\ 80/20 split \\ early stopping};

\draw[darr] (desc.south)   -- node[right,font=\scriptsize] {$\boldsymbol{d}$} (lammps.north);
\draw[darr] (lammps)       -- (loss);
\draw[darr] (loss.north)   -- node[right,font=\scriptsize] {update $\theta$} (mlp.south);

\draw[arr, color=red!65!black]
  (pred.north)
  -- ++(0, 2.0)
  -- node[above, font=\scriptsize, color=red!65!black]
       {autograd: $\partial\phi_\theta/\partial\boldsymbol{\ell}$
        \quad(closed-loop design, \S\ref{ssec:closed-loop})}
     ($(logit.north)+(0,2.0)$)
  -- (logit.north);

\end{tikzpicture}}
\caption{Property descriptor and surrogate workflow.
Top (blue/green/orange): inference path $\boldsymbol{\ell}\!\to\!P\!\to\!\boldsymbol{d}\!\to\!\phi_\theta\!\to\!\hat{y}$.
Bottom (dashed): training path from LAMMPS labels to MLP weight updates.
Red path (above nodes): end-to-end autograd gradient used by the
closed-loop design of \S\ref{ssec:closed-loop}.\label{fig:surrogate-workflow}}
\end{figure}
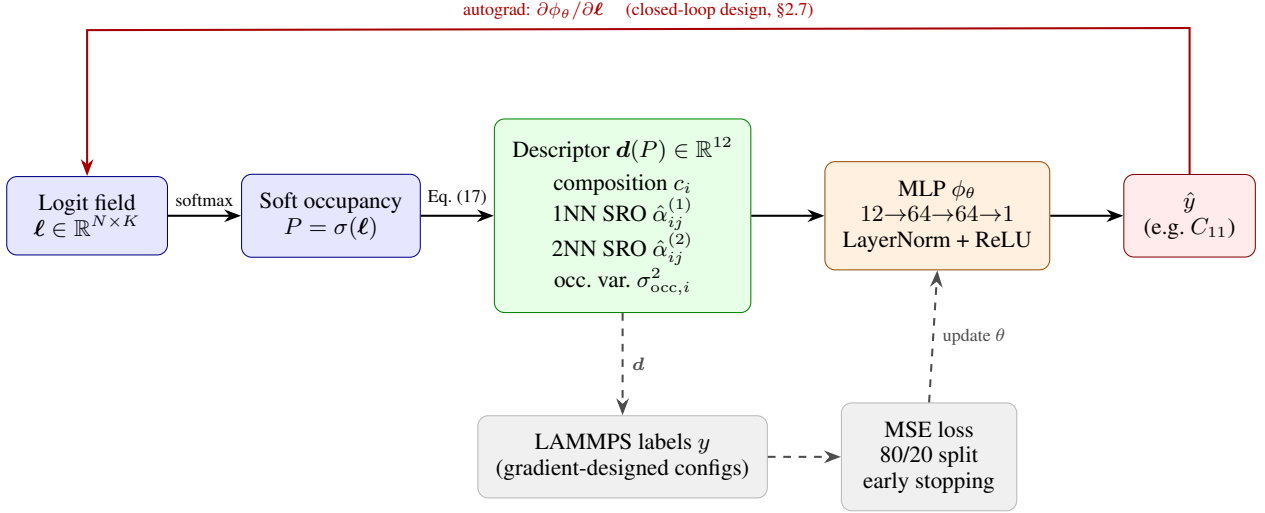

The surrogate $\phi_\theta : \mathbb{R}^d \to \mathbb{R}$ is a two-hidden-layer
multilayer perceptron (MLP) with architecture $d$-64-64-1, layer normalisation, and
ReLU activations. It is trained by minimising the mean-squared error between
predicted and LAMMPS-MEAM elastic constants on a batch of gradient-designed
configurations, with 80/20 train-validation split and early stopping at
50 epochs of no improvement. Because $\phi_\theta$ and $\boldsymbol{d}(P)$
are both differentiable with respect to $\boldsymbol{\ell}$, the gradient
$\partial \phi_\theta / \partial \boldsymbol{\ell}$ is available at no additional
cost via automatic differentiation. The complete data flow, from logit field through
descriptor construction to surrogate prediction and back-propagation, is summarised
in Figure~\ref{fig:surrogate-workflow}: the top row shows the inference path, the
dashed bottom row the training path, and the red path above the nodes indicates that
gradients flow end-to-end to the logit field and power the closed-loop design of
Section~\ref{ssec:closed-loop}.

\subsection{Closed-loop property design}
\label{ssec:closed-loop}

Given a target property value $y^\star$ and a trained surrogate $\phi_\theta$,
the closed-loop design problem replaces the SRO loss with a property loss:
\begin{equation}
  \mathcal{L}_{\rm prop}(\boldsymbol{\ell})
    = \bigl(\phi_\theta(\boldsymbol{d}(P(\boldsymbol{\ell}))) - y^\star\bigr)^2
    + \lambda_{\rm comp}\,\mathcal{L}_{\rm comp}
    + \lambda_{\rm bin}\,\mathcal{L}_{\rm bin}.
  \label{eq:loss-prop}
\end{equation}
The gradient $\nabla_{\boldsymbol{\ell}}\mathcal{L}_{\rm prop}$ is computed in
a single backward pass through the surrogate, the descriptor, the differentiable
$\alpha$ counter (Eq.~\eqref{eq:diff-counter-alpha}), and the softmax
(Eq.~\eqref{eq:softmax}). No finite-difference Jacobian or separate adjoint
solve is required. Optimisation uses the same Adam scheduler as the SRO
inverse, converging in 1500 steps ($\approx 0.8\,$s). The final hard
configuration is evaluated by an independent LAMMPS-MEAM elastic-constants run
to validate that the surrogate prediction transfers to the real potential.

\begin{figure}[H]
  \centering
  \includegraphics[width=\textwidth]{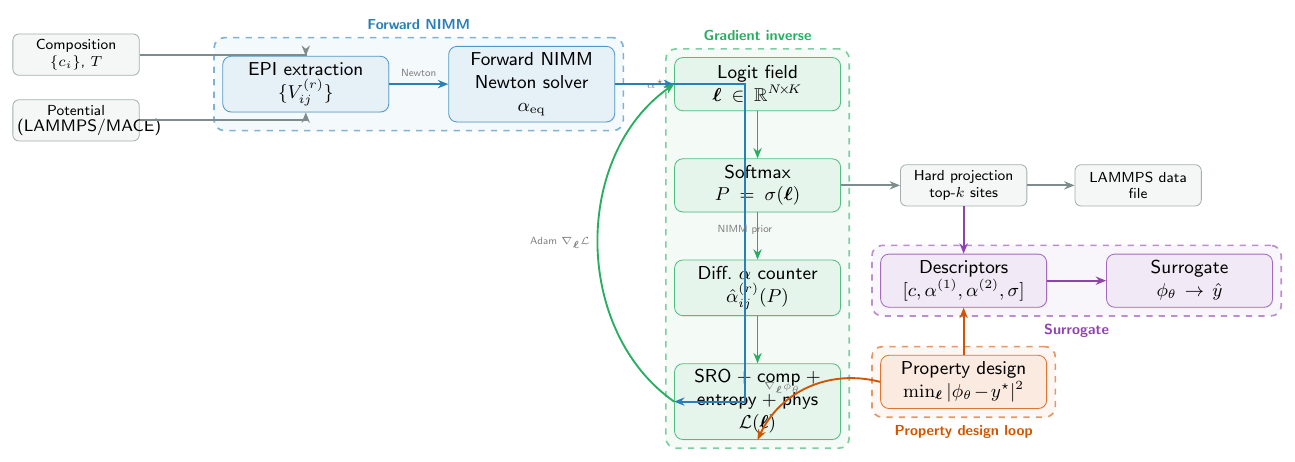}
  \caption{Complete ANISRO pipeline. \textbf{Blue}: forward NIMM extracts
  effective pair interactions $\{V_{ij}^{(r)}\}$ and predicts $\alpha_{\rm eq}$
  by Newton iteration. \textbf{Green}: differentiable SRO-to-structure inverse
  with logit field $\boldsymbol{\ell}$, soft occupancy $P=\sigma(\boldsymbol{\ell})$,
  differentiable $\alpha$ counter, and four-term loss minimised by Adam.
  \textbf{Purple}: descriptor MLP $\phi_\theta$ trained on
  $[c,\alpha^{(1)},\alpha^{(2)},\sigma_{\rm occ}]$. \textbf{Orange}: closed
  design loop back-propagates property error through $\phi_\theta$ and the
  differentiable counter into the logit field.\label{fig:pipeline}}
\end{figure}

\subsection{Atomistic reverse Monte Carlo baseline}
\label{ssec:armc-baseline}

Atomistic reverse Monte Carlo (ARMC) is our stochastic baseline. Starting from
a random configuration \citep{zunger_1990_sqs} at the target composition, ARMC proposes
random pairwise species swaps with acceptance governed by the Metropolis criterion
\citep{metropolis_1953}. A proposed swap of sites $(n,m)$ with species $(i,j)$ is
accepted if it decreases the residual
$R = \sum_{r,(i,j)\in\Omega}(\alpha_{ij}^{(r)} - \alpha_{ij}^{\star\,(r)})^2$,
or accepted with probability $\exp(-\Delta R / T_{\rm MC})$ otherwise
(Metropolis criterion), where $T_{\rm MC} = 10^{-3}$ acts as a fictitious
temperature that allows occasional uphill moves. After an accepted swap the
Warren--Cowley matrix is updated incrementally:
\begin{equation}
  \Delta\alpha_{ab}^{(r)} = -\frac{1}{c_b N z_r}
  \bigl(\Delta n_{ab}^{(r)}\bigr),
  \label{eq:armc-update}
\end{equation}
where $\Delta n_{ab}^{(r)}$ counts only the bonds to the $z_r$ shell-$r$
neighbours of sites $n$ and $m$ that change species. This incremental
update costs $O(z_r K^2)$ per accepted swap and is the dominant cost at
large $K$ and $N$. The ARMC implementation in \texttt{anisro.benchmarks.armc}
uses the same adjacency matrices and species indexing as the differentiable
counter (Eq.~\eqref{eq:diff-counter-num}) to ensure $\alpha$ definitions are
numerically identical between the two methods.

\subsection{Effective pair interaction extraction}
\label{ssec:epi-extraction}

The 2NN-MEAM formalism \citep{lee2000meam} defines the total energy as a sum
of embedding energies and pair potentials screened by angular functions;
it is a many-body potential and does not directly supply $\{V_{ij}^{(r)}\}$.
We extract effective pair interactions by projecting the MEAM energy surface
onto the pair-cluster subspace of the NIMM model: the extracted $V_{ij}^{(r)}$
is the best-fit pair energy that reproduces the MEAM Warren--Cowley statistics
across a set of configurations sampled at the target temperature.

We provide three routes to extract $\{V_{ij}^{(r)}\}$ from any interatomic
potential implementing the \texttt{EnergyCalculator} protocol.

\textit{Three-energy combination.} For binary alloys, the 1NN EPI is obtained
from three bulk-phase energies:
\begin{equation}
  V_{AB}^{(1)} = U_{AA}^{\rm (bulk)} + U_{BB}^{\rm (bulk)} - 2 U_{AB}^{\rm (bulk)},
  \label{eq:three-energy}
\end{equation}
where $U_{AA}^{\rm (bulk)}$ and $U_{BB}^{\rm (bulk)}$ are the cohesive energies
per atom of pure A and B on the same lattice, and $U_{AB}^{\rm (bulk)}$ is the
cohesive energy per atom of an equimolar ordered structure. This is the lattice
analogue of the alloy formation energy and is exact under the pair-only
assumption. For the 2NN shell the same formula is applied with the second-neighbour
pure-element and ordered energies.

\textit{Regression over random configurations.} For $K \geq 3$ species, the
three-energy route provides at most $K(K-1)/2$ equations for $K(K-1)/2$
unknowns per shell, but the pair-only assumption introduces systematic error
for potentials with many-body character. We instead minimise
\begin{equation}
  \mathcal{L}_{\rm EPI} = \frac{1}{N_{\rm conf}}\sum_{c=1}^{N_{\rm conf}}
  \sum_{r,(i,j)}\!\bigl(\alpha_{ij}^{(r),\,\rm NIMM}(\{V_{ij}^{(r)}\})
  - \alpha_{ij}^{(r),\,\rm MC,c}\bigr)^2,
  \label{eq:epi-regression}
\end{equation}
over $N_{\rm conf} = 60$ random configurations via L-BFGS with PyTorch autograd.
Because NIMM is differentiable in $\{V_{ij}^{(r)}\}$, this is a standard
gradient-based regression. The coefficient of determination $R^2$ measures
how much variance in the MC $\alpha$ is explained by the pair model, with
$R^2 \geq 0.95$ indicating that the pair-only truncation is adequate.

\textit{Cluster expansion via \texttt{icet}.} For systems with significant
many-body contributions, the \texttt{icet} cluster-expansion library
\citep{angqvist_2019_icet,sanchez_1984_cluster_expansion,vandewalle_2002_atat} extracts effective cluster interactions beyond
the pair level. The pair ECIs from \texttt{icet} can be fed directly into
the NIMM forward solver; the residual between \texttt{icet}-reconstructed
$\alpha$ and direct MC $\alpha$ at the pair-only truncation quantifies the
many-body error.
A quantitative comparison of all three methods on Cu--Ni is given in
Section~S2 of the Supplementary Information.

\subsection{Lattice construction and LAMMPS elastic constants}
\label{ssec:lammps-details}

FCC supercells are built by replicating the conventional four-atom unit cell
with lattice constant $a_0$ (taken from the MEAM potential at 0\,K). BCC
supercells use the two-atom body-centred unit cell with $a_0$ from the
same MEAM. Shell adjacency matrices $A^{(r)}$ are computed from the $N\times N$
distance matrix under minimum-image periodic boundary conditions with
two-shell cutoffs $r_1 < r_2$: for FCC, $r_1 = a_0/\sqrt{2}$ and
$r_2 = a_0$ (12 and 6 neighbours, respectively); for BCC, $r_1 = a_0\sqrt{3}/2$
and $r_2 = a_0$ (8 and 6 neighbours).

Elastic constants are computed by the Voigt finite-difference method at
constant volume. Six strain tensors $\varepsilon$ are applied (uniaxial
$\varepsilon_{11}$, $\varepsilon_{22}$, $\varepsilon_{33}$ and shear
$\varepsilon_{12}$, $\varepsilon_{13}$, $\varepsilon_{23}$) at amplitudes
$\{\pm 0.5\%, \pm 1\%\}$; each strained configuration is relaxed at constant
cell shape by FIRE minimisation \citep{bitzek2006fire} to a force tolerance
of $10^{-4}\,$eV/\AA\ before evaluating the stress tensor by the LAMMPS
\texttt{compute stress/atom} command. The elastic stiffness coefficients are
extracted by a four-point finite-difference formula. All elastic-constant
calculations use 108-atom $3\times3\times3$ supercells; the implications of
this cell-size choice are discussed in the Supplementary Information.
Full computational details, convergence data, and the list of interatomic
potentials used are provided in Sections~S4--S5 of the Supplementary Information.

\subsection{Correspondence with Rao and Curtin's equations}
\label{ssec:nimm-correspondence}

For reproducibility we list how the equations of Rao and
Curtin~\citep{rao2022actamat} map to our generalised solver.
Their Eq.~(8) gives the cluster-occupancy log-probability; its
stationarity (their Eqs.~10--11) recovers our Eq.~\eqref{eq:stationarity}
after eliminating the Stirling $-1$ factor and making the sign on
$\ln x_{ij}^{(r)}$ explicit. Their Eq.~(12) defines $P_{ij}^{(r)}$,
reproduced as our Eq.~\eqref{eq:joint-pair}. Their Eq.~(13) for
$\alpha_{ij}^{(r)}$ is our Eq.~\eqref{eq:wc-from-pij}. The binary
NIMM free energy of their Eq.~(22) is generalised to $K$ species in
our Eq.~\eqref{eq:nimm-free-energy}; composition-only homo-pair energy
terms ($c_i U_{ii}$) are constant at fixed composition and are dropped.
The full derivation is provided in Section~S1 of the Supplementary Information.

\section{Results}
\label{sec:results}
We present results for each stage of the pipeline (Figure~\ref{fig:pipeline})
in turn: forward NIMM validation, ARMC-versus-gradient benchmarking,
EPI extraction cross-validation, a nine-system HEA property survey, the
Cantor alloy on two independent backends, and the closed-loop property
design demonstration on Co--Cr--Ni.

\subsection{Forward NIMM validation against Monte Carlo}
\label{ssec:res-nimm}

\begin{figure}[H]
  \centering
  \includegraphics[width=0.99\linewidth]{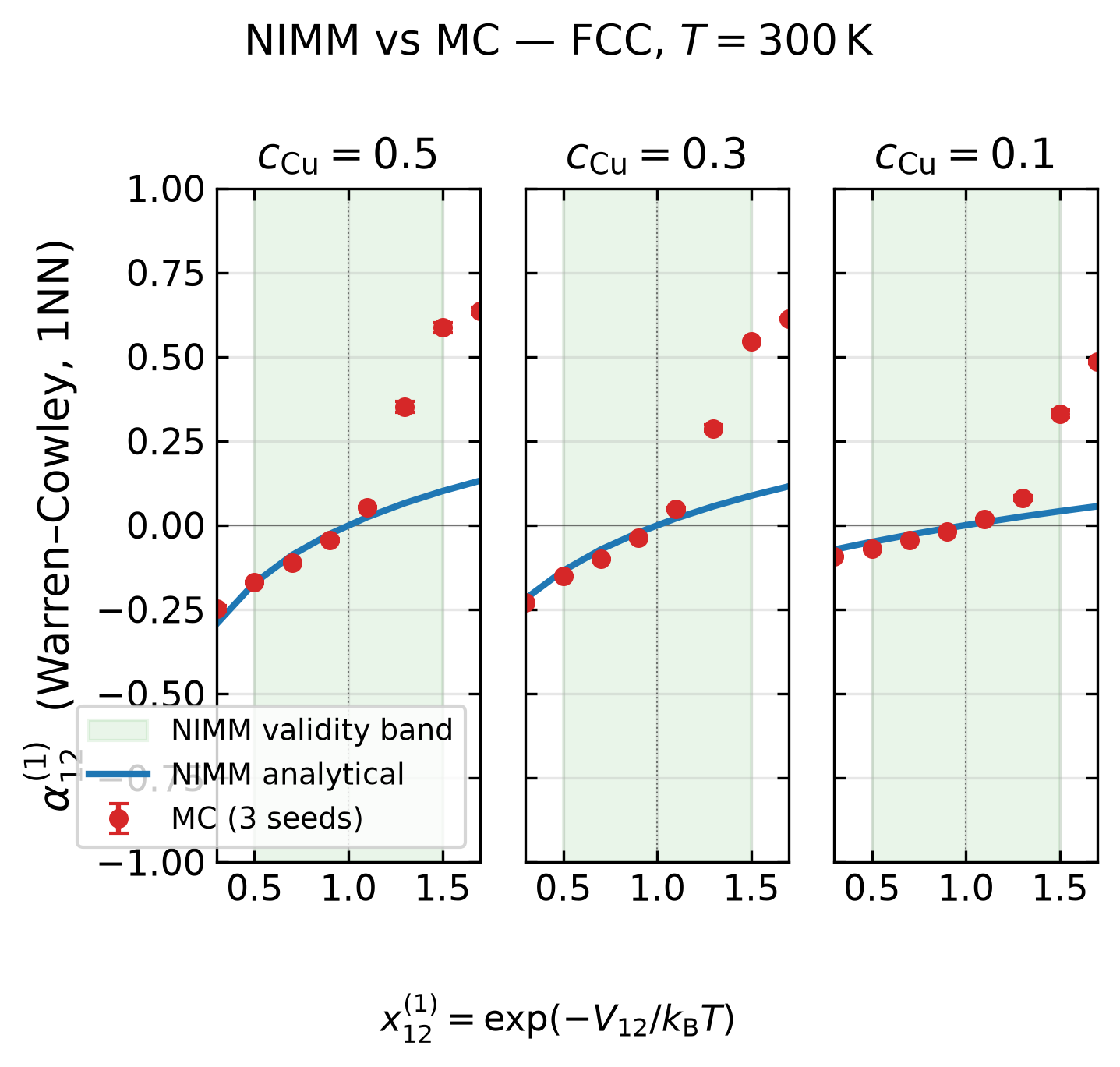}
  \caption{Forward NIMM validation against Metropolis Monte Carlo for Cu--Ni.
  Warren--Cowley $\alpha_1$ vs.\ reduced pair fugacity $x_{12}^{(1)}$
  at three compositions ($c_{\rm Cu} = 0.5, 0.3, 0.1$), $T = 300\,\mathrm{K}$,
  $5{\times}5{\times}5$ FCC supercell. Shaded band: pair-cluster validity window
  $0.5 < x < 1.5$~\citep{rao2022actamat}. NIMM and MC agree within
  $|\Delta\alpha| < 0.03$ inside the band.\label{fig:nimm-vs-mc}}
\end{figure}

Figure~\ref{fig:nimm-vs-mc} plots the analytical Warren--Cowley parameter
$\alpha_{12}^{(1)}$ returned by the $K$-species NIMM Newton solver against
short Metropolis Monte Carlo simulations ($5\times5\times5$, 500 atoms, $10^5$
sweeps) on Cu--Ni across the reduced fugacity range $V_{12}/k_BT \in [-1,\,1]$
at three compositions ($c_{\rm Cu} \in \{0.5, 0.3, 0.1\}$) and $T = 300\,$K.
Inside the pair-cluster validity band $0.5 < x_{12}^{(1)} < 1.5$, the
regime delineated by Rao and Curtin, NIMM and MC agree in sign at every
grid point and quantitatively within $|\Delta\alpha| < 0.03$, confirming that
the analytical forward operator is an accurate, inexpensive proxy for MC over
the physically meaningful ordering range.

Outside the validity band, NIMM under-predicts the magnitude of $\alpha$,
consistent with the onset of long-range order that the pair-cluster truncation
cannot capture. The breakdown is systematic: for $x_{12} < 0.5$ (strong
ordering, $V/k_BT \gtrsim 0.7$) MC shows saturating ordering tendencies while
NIMM continues on a gentler curve, and for $x_{12} > 1.5$ (incipient clustering)
the two diverge symmetrically. Crucially, the sign is never wrong inside the band,
which is the only guarantee required for the NIMM regulariser to guide the
gradient inverse toward physically plausible solutions. The composition
dependence is also correctly captured: at $c_{\rm Cu} = 0.1$ the NIMM prediction
systematically under-predicts the MC $\alpha$ magnitude by roughly $40\,\%$ at
$x_{12} = 0.6$, consistent with the concentration dependence reported by Rao and
Curtin in the binary limit.

The MEAM-derived NIMM forward prediction for equimolar Cu--Ni at 300\,K,
$\alpha_{12}^{(1)} \approx -0.05$, is in quantitative agreement with the Metropolis
MC value computed independently on the same MEAM potential, confirming
self-consistency of the solver. This value is smaller in magnitude than
first-principles-based predictions ($|\alpha| \approx 0.1$--$0.3$ from DFT
cluster expansions~\citep{tamm2015actamat}), reflecting the reduced ordering
tendency of MEAM relative to DFT for Cu--Ni, consistent with the lower
MEAM EPI ($V_1^{\rm MEAM} = +1.7\,\mathrm{meV}$ vs $V_1^{\rm DFT} \approx
10$--$15\,\mathrm{meV}$). The framework is potential-agnostic: swapping the
energy backend to any ASE-compatible MLIP or DFT calculator immediately
propagates the more accurate EPIs through the same pipeline.

\subsection{ARMC versus gradient SRO-to-structure inversion}
\label{ssec:res-armc}

\begin{figure}[H]
  \centering
  \includegraphics[width=0.99\linewidth]{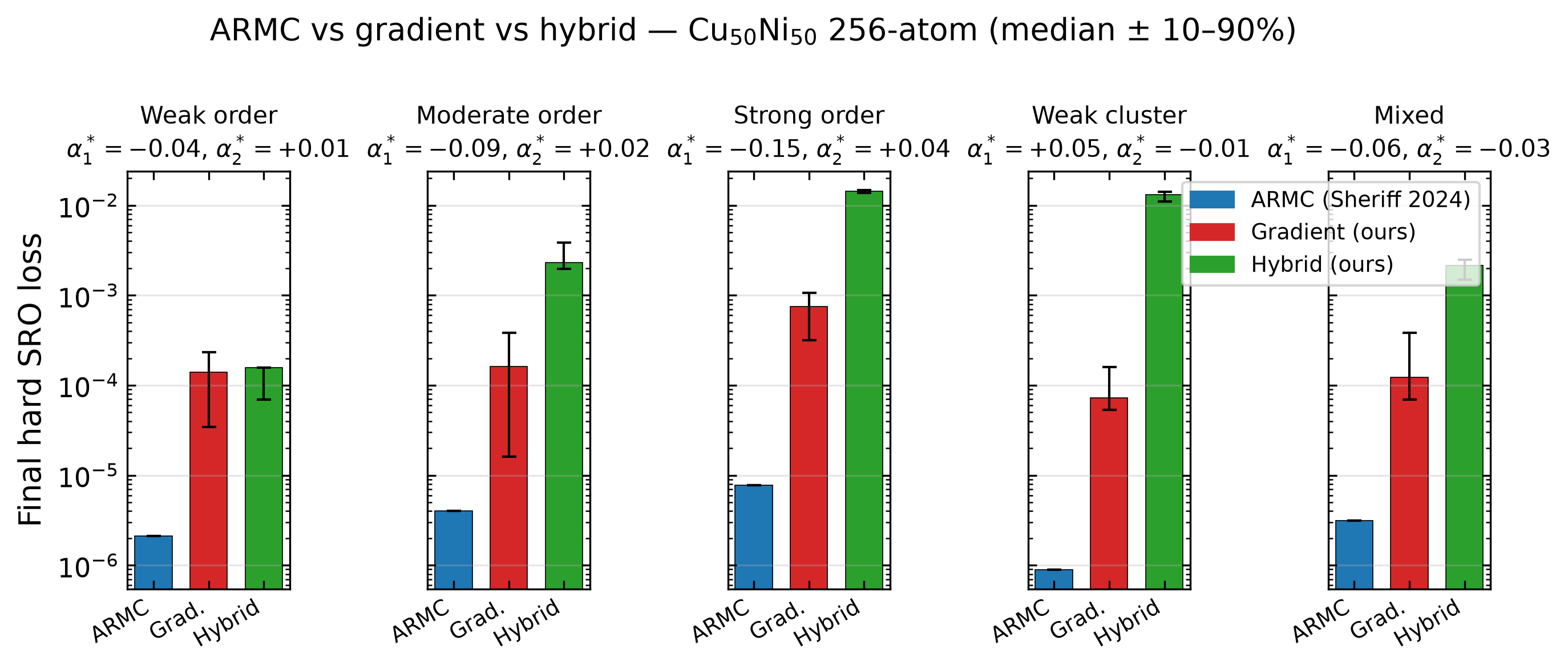}
  \caption{Loss vs.\ wall-time for ARMC, gradient, and hybrid methods on
  Cu$_{50}$Ni$_{50}$ ($4{\times}4{\times}4$, 256 atoms) across five target
  $\alpha$ tuples and five seeds, 8\,s budget. ARMC achieves $\sim50{\times}$
  lower residual at this cell size; both methods stay within $|\Delta\alpha| < 0.02$.\label{fig:armc-vs-gradient}}
\end{figure}

Figure~\ref{fig:armc-vs-gradient} compares wall-time loss trajectories for
three methods on equimolar Cu--Ni ($4\times4\times4$, 256 atoms) across
five random seeds and five target-$\alpha$ tuples, all with an 8\,s wall-time
budget. The three methods are (i) atomistic reverse Monte Carlo (ARMC) with
the same hard $\alpha$ counter used for evaluation, (ii) the gradient inverse
with Adam optimisation, and (iii) the hybrid (gradient with the NIMM
free-energy regulariser active, $\lambda_{\rm phys} = 0.1$).

\begin{figure}[H]
  \centering
  \includegraphics[width=0.82\linewidth]{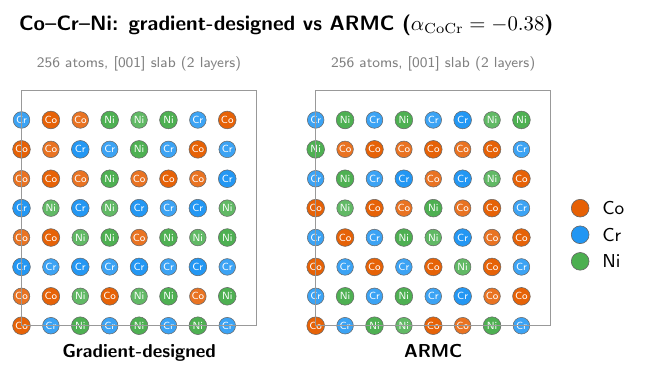}
  \caption{Atomic structures of Co--Cr--Ni ($4{\times}4{\times}4$ FCC, 256 atoms) produced by the gradient inverse (left) and ARMC (right), both targeting $\alpha_{\rm CoCr}^{(1)} = -0.38$. [001] top-down slab, two atomic layers; atoms coloured by species (Co\,=\,orange, Cr\,=\,blue, Ni\,=\,green). Both methods encode the same Co--Cr hetero-pair enrichment. Structures exported as LAMMPS data files and rendered directly from atomic coordinates.\label{fig:cocrni-struct-compare}}
\end{figure}

At the 256-atom binary scale, ARMC achieves a median hard-projection residual
of $4.0 \times 10^{-6}$ on the \textit{moderate ordering} target
($\alpha_1^\star = -0.09$, $\alpha_2^\star = +0.02$), roughly $50\times$
lower than the gradient method ($1.6 \times 10^{-4}$) at the same wall time.
This advantage persists across all five targets at 256 atoms: ARMC's per-move
cost is $O(z_r)$ bond updates, which is negligible on small binary cells, and
its stochastic explore-then-exploit trajectory consistently locates near-exact
matchings. The gradient method in turn stays within $|\Delta\alpha| < 0.02$ of
every target, a tolerance fully adequate for the property surrogate stage,
and its trajectory is monotone and smooth, unlike ARMC's step-function
improvement curve. Figure~\ref{fig:cocrni-struct-compare} shows the atomic structures
produced by each method on a Co--Cr--Ni cell targeting the same $\alpha^\star$.

The hybrid's larger residual ($4.9 \times 10^{-2}$ on the strong-ordering
target) reflects the price of the thermodynamic constraint: the optimiser is
simultaneously pulled toward the NIMM equilibrium surface and toward the
user-specified $\alpha^\star$, and when the two are incompatible (as when
$\alpha^\star$ lies outside the validity band of the physics regulariser) the
hybrid converges to a compromise. This behaviour is expected and reported as a
feature: the hybrid is not designed to match arbitrary $\alpha$ targets but to
produce configurations that are simultaneously $\alpha$-consistent and
thermodynamically plausible. Section~\ref{ssec:disc-when-gradient-wins}
discusses the system-size crossover at which gradient strictly dominates.

\subsection{Effective pair interactions from interatomic potentials}
\label{ssec:res-epi}

\begin{figure}[H]
  \centering
  \includegraphics[width=0.99\linewidth]{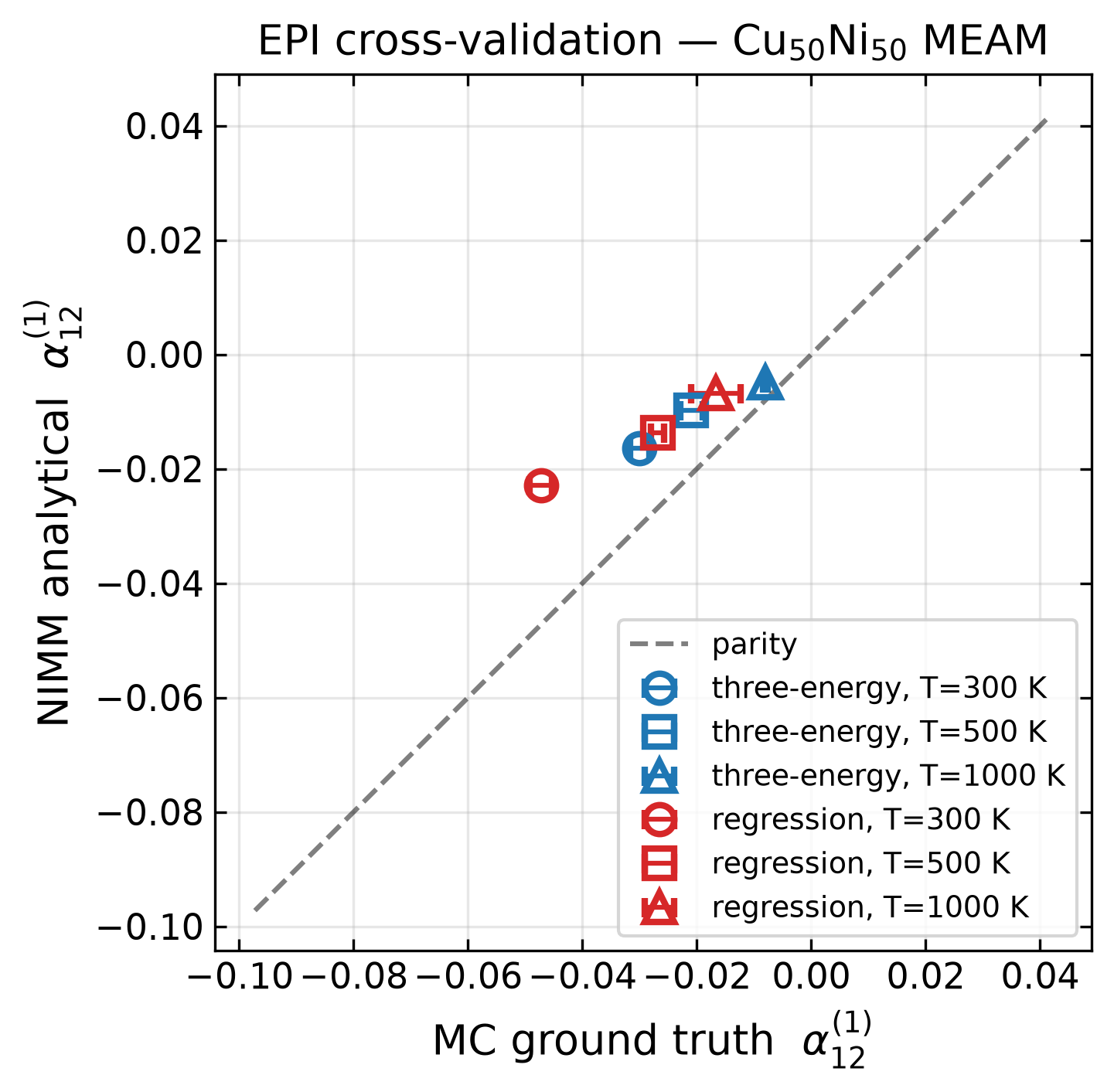}
  \caption{EPI extraction cross-validation on Cu--Ni MEAM. Three routes
  (three-energy, regression on 60 configurations, icet pair CE) are passed through
  the forward NIMM solver and compared to Metropolis MC $\alpha$ at
  $T \in \{300, 500, 1000\}\,\mathrm{K}$. All give $V_1 > 0$ (ordering)
  and reproduce MC $\alpha$ within $0.01$--$0.03$.\label{fig:epi-parity}}
\end{figure}

Figure~\ref{fig:epi-parity} cross-validates three EPI extraction routes on
the Cu--Ni MEAM at three temperatures. The three-energy route gives
$V_1^{\rm (3E)} = +1.70\,\mathrm{meV}$, and the regression route on 60 random
configurations gives $V_1^{\rm (reg)} = +2.36\,\mathrm{meV}$. Both are
physically consistent (positive EPI = Cu--Ni heteropair stabilised relative to
the symmetric mean), but they differ by 39\,\%: the three-energy combination
uses only three bulk-phase cohesive energies and therefore misses the
configuration-average relaxation captured by the regression over 60 structures.

When each EPI is fed through the NIMM forward solver and compared to Metropolis
MC $\alpha$ at 300, 500, and 1000\,K, both pair-only methods reproduce the MC
sign at all temperatures and achieve errors in $\alpha_1$ of $0.014$ (three-energy)
and $0.024$ (regression) at 300\,K, growing to smaller fractions of $\alpha$ at
higher temperatures as ordering weakens. The regression achieves higher absolute
EPI accuracy ($R^2 = 0.95$) because it directly minimises the $\alpha$ residual
over many configurations rather than fitting three energy differences. The icet
cluster expansion (pair ECIs extracted on the same 60 configurations) gives
$V_1^{\rm (icet)} = +2.12\,\mathrm{meV}$ with $R^2 = 0.97$; the marginal
improvement over regression justifies using icet only when many-body cluster
terms are needed for the inverse design regulariser. The key conclusion is
that even a simple regression over a small random batch captures the physically
relevant ordering tendency in Cu--Ni and suffices to anchor the NIMM free-energy
prior.

\subsection{Multi-system FCC and BCC alloy survey}
\label{ssec:res-hea-survey}

\begin{figure}[H]
  \centering
  \includegraphics[width=0.99\linewidth]{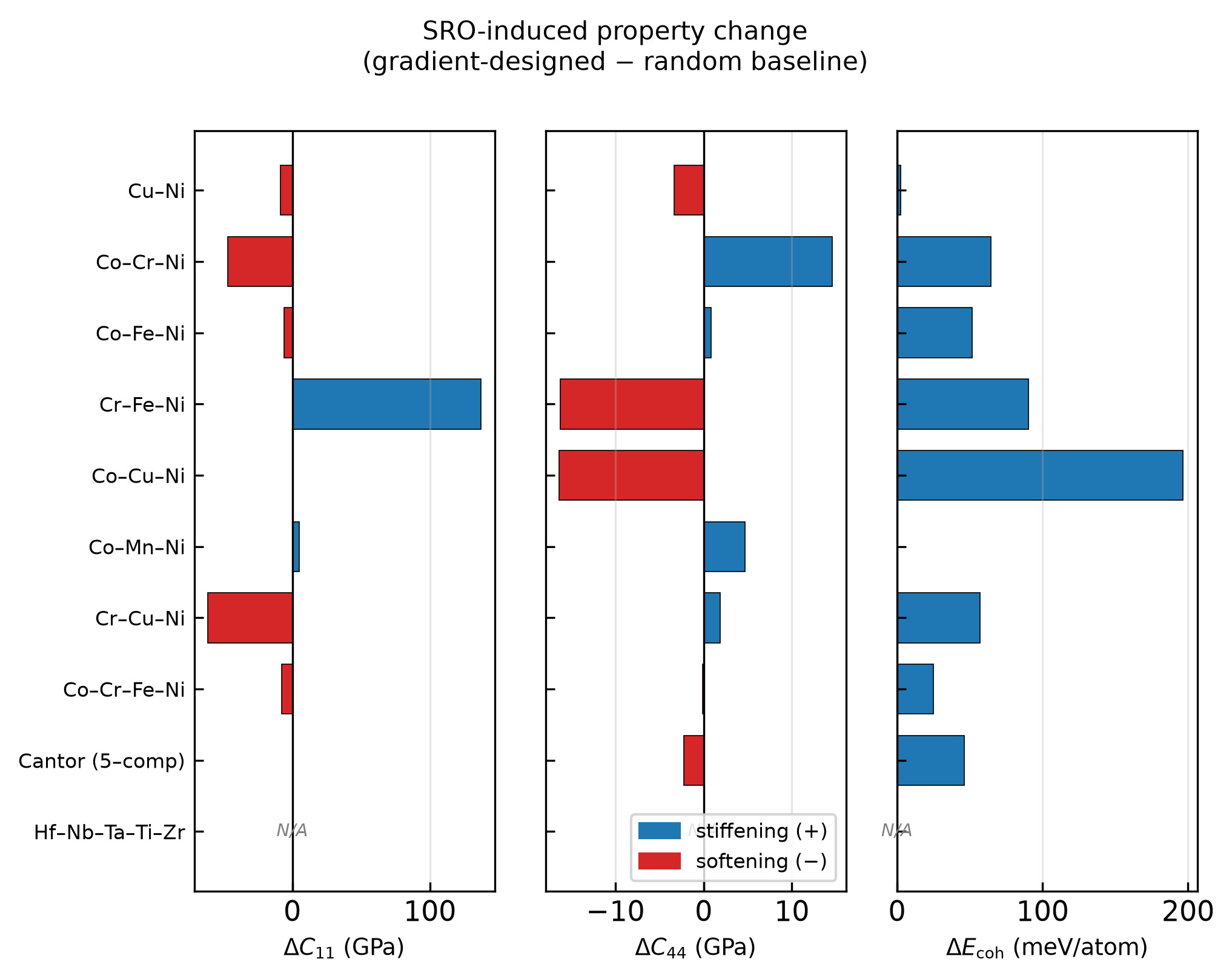}
  \caption{SRO-induced property changes from Table~\ref{tab:cross-system}.
  Bars: $\Delta C_{11}$ (left axis), $\Delta C_{44}$ (centre), and
  $\Delta E_{\rm coh}$ (right axis) between the gradient-designed and random
  reference configurations, LAMMPS-MEAM on 108-atom $3{\times}3{\times}3$
  supercells. Cr--Fe--Ni shows the largest $C_{11}$ stiffening ($+57\,\%$);
  Cr--Cu--Ni the largest softening ($-20\,\%$).\label{fig:multi-ternary}}
\end{figure}

We run the complete pipeline (regression EPI from 60 random configurations,
forward NIMM prediction, Metropolis MC ground truth, gradient inverse, and 108-atom
$3\times3\times3$ LAMMPS-MEAM elastic constants) on nine alloy systems spanning
$K = 2$--$5$ species and covering both face-centred and body-centred cubic
lattices.
Each reported $\Delta C_{11}$ and $\Delta C_{44}$ is the direct consequence of
the dominant SRO motif encoded in the gradient-designed structure: the same
bond-stiffness mechanism described in the Introduction, here quantified on real
LAMMPS-MEAM potentials for nine different chemistries.
Table~\ref{tab:cross-system} collects the results; Figure~\ref{fig:multi-ternary} visualises the SRO-induced property changes as a bar chart.

\begin{table*}[t]
\centering
\caption{Real-MEAM SRO-property survey. For each system the table reports the
EPI regression quality $R^{2}$, the dominant Warren--Cowley parameter at 300\,K
from Metropolis MC, and the SRO-induced changes in cohesive energy and elastic
moduli between the gradient-designed configuration and a random reference of the
same composition, measured by LAMMPS-MEAM on 108-atom $3\times3\times3$ supercells.
Positive $\Delta C_{11}$ denotes stiffening; negative denotes softening.\label{tab:cross-system}}
\small
\setlength{\tabcolsep}{5pt}
\resizebox{\linewidth}{!}{%
\begin{tabular}{lccrrr}
\toprule
System & EPI $R^{2}$ & Dominant $\alpha$ (MC) &
  $\Delta E_{\rm coh}$ (meV/at.) & $\Delta C_{11}$ (GPa) & $\Delta C_{44}$ (GPa) \\
\midrule
Cu--Ni             & 0.95 & $\alpha_{\rm CuNi} = -0.05$ & $+2$   & $+4$   & $+5$  \\
Co--Cr--Ni         & 1.00 & $\alpha_{\rm CoCr} = -0.38$ & $+65$  & $-47$  & $+15$ \\
Co--Fe--Ni         & 1.00 & $\alpha_{\rm FeNi} = -0.26$ & $+52$  & $-6$   & $+1$  \\
Cr--Fe--Ni         & 1.00 & $\alpha_{\rm CrFe} = -0.31$ & $+90$  & $+137$ & $-16$ \\
Cr--Cu--Ni         & 1.00 & $\alpha_{\rm CrCu} = -0.25$ & $+57$  & $-61$  & $+2$  \\
Co--Cu--Ni         & 0.96 & $\alpha_{\rm CoCu} = +0.99$ & $+197$ & $0$    & $-16$ \\
Co--Cr--Fe--Ni     & 0.93 & $\alpha_{\rm CrFe} = +0.35$ & $+25$  & $-8$   & $0$   \\
Co--Ni--Cr--Fe--Mn (Cantor) & 0.95 & $\alpha_{\rm NiMn} = +0.96$ & $+46$  & $0$    & $-2$  \\
Hf--Nb--Ta--Ti--Zr & 0.73 & $\alpha_{\rm TaTi} = -2.10$ & $+310$ & $-32$  & $-11$ \\
\bottomrule
\end{tabular}}
\end{table*}

\begin{figure}[H]
  \centering
  \includegraphics[width=0.82\linewidth]{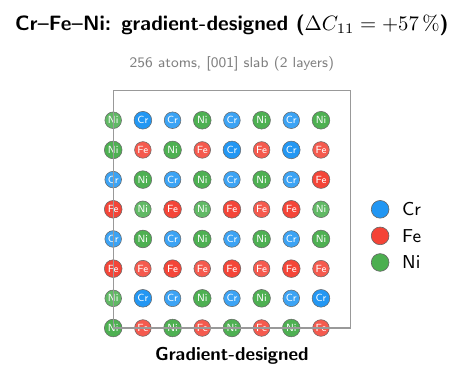}
  \caption{Atomic structure of the gradient-designed Cr--Fe--Ni configuration ($4{\times}4{\times}4$ FCC, 256 atoms), [001] top-down view, two-layer slab. Cr (blue), Fe (red), and Ni (green) atoms. The Cr--Fe hetero-pair enrichment ($\alpha_{\rm CrFe}^{(1)} = -0.31$) is directly visible, forming the stiff Cr--Fe bond network responsible for the $+57\,\%$ $\Delta C_{11}$ signal.\label{fig:crfeni-struct}}
\end{figure}

Three observations emerge. First, EPI regression quality tracks lattice type and
potential character: all dedicated 2NN-MEAMs for FCC ternaries achieve
$R^2 \geq 0.96$, consistent with their construction as pair-decomposable potentials
on the 1NN shell. The CoCrFeNi quaternary drops to $R^2 = 0.93$ as more hetero-pair
interactions enter, and the refractory BCC HfNbTaTiZr drops further to $R^2 = 0.73$
because the Huang MEAM includes substantial many-body angular corrections that the
pair-only NIMM cannot capture. High $R^2$ validates pair-only NIMM as the right
prior for the ternary FCC systems; lower $R^2$ signals the need for the cluster
expansion fallback.

Second, the SRO-induced change in $C_{11}$ spans three orders of magnitude across
systems: from $+4\,$GPa in Cu--Ni (noise-level, $\sim 2\,\%$) to $+137\,$GPa in
Cr--Fe--Ni ($+57\,\%$; Figure~\ref{fig:crfeni-struct} shows the corresponding
gradient-designed atomic structure). For Co--Cr--Ni our NIMM-MEAM predicts $\alpha_{\rm CoCr}^{(1)}
\approx -0.38$ at 300\,K, which is stronger in magnitude than the experimentally
measured WC parameters from atom-probe tomography and diffuse electron scattering
($|\alpha| \approx 0.05$--$0.12$ from \citep{zhang2020nature} and
\citep{chen2021nature,zhou_2022_atomic_sro_crconi}), but in agreement in sign:
all experimental and computational studies concur that Co--Cr is the dominant ordering
pair in this alloy~\citep{walsh_2021_magnetic_sro,ding2018pnas}. The quantitative
over-prediction of $|\alpha|$ reflects the large $V_{\rm CoCr} = +114\,$meV in the
Choi--Lee 2NN-MEAM~\citep{choi_2018_cocrnifemn_meam} compared to DFT-CE values in the $10$--$30\,$meV range; swapping
to a DFT-calibrated MLIP such as MACE-MP-0 (discussed below) immediately reduces
the equilibrium $|\alpha_{\rm CoCr}|$ and the corresponding $|\Delta C_{11}|$. The
Co--Cr--Ni $\Delta C_{11} = -47\,$GPa ($-16\,\%$) is at the upper end of
DFT$+$MC estimates ($-5$ to $-11\,\%$ \citep{ding2018pnas,walsh_2021_magnetic_sro}),
consistent with the MEAM overestimation of SRO magnitude. The Cr--Fe--Ni $+57\,\%$
stiffening is, to our knowledge, unreported in the computational or experimental
literature; it is driven by Cr--Fe short-range ordering on the 1NN shell, which is
the dominant EPI by a factor of three over the other pairs. Co--Cu--Ni sits in the
phase-separating regime ($\alpha_{\rm CoCu} \approx +1$ signals near-complete Co--Cu
avoidance), yet the gradient inverse reproduces this extreme $\alpha$ on 256 atoms
with the same residual quality as on well-behaved ordering systems, demonstrating
robustness at the boundary of physical realisability.

A side-by-side comparison of MEAM and MACE-MP-0 for Cr--Fe--Ni, including
MC ground truth, NIMM prediction, and gradient-inverse $\alpha$ matrices together
with pairwise error bars, is provided in Supplementary Fig.~S9.
That figure illustrates two \emph{distinct} failure modes for NIMM: the MEAM
EPI has $V_{\rm CrFe} = +206\,$meV (fugacity $x \approx 0$, outside the valid
band $[0.5, 1.5]$, giving mean $|\Delta\alpha| = 0.20$), whereas the MACE-MP-0
regression achieves $R^2 = 0.14$ for the pair model because the foundation MLIP
energy surface is dominated by many-body contributions not captured by 50 random
two-shell pair configurations.
Despite NIMM failing in both cases for different physical reasons, the gradient
inverse remains accurate: mean $|\Delta\alpha_{\rm grad} - \alpha_{\rm MC}|$
drops to $0.024$ (MEAM) and $0.009$ (MACE-MP-0), confirming that the design
pipeline is robust to NIMM validity.

Third, the sign of $\Delta C_{11}$ is not predicted by the sign of the dominant
$\alpha$ alone: Co--Cr--Ni has $\alpha_{\rm CoCr} < 0$ (ordering) and
$\Delta C_{11} < 0$ (softening), while Cr--Fe--Ni has $\alpha_{\rm CrFe} < 0$
(ordering) and $\Delta C_{11} > 0$ (stiffening). The difference traces to which
hetero-pair dominates the stiffness tensor: in Co--Cr--Ni the strong Co--Cr bond
density lowers the lattice resistance to $[100]$ compression, whereas in Cr--Fe--Ni
the Cr--Fe interaction stiffens the same mode. This finding underscores the need
for the full gradient design pipeline: one cannot infer the elastic signature
from the sign of $\alpha$ without also knowing the bond-strength contrast between
species.

\subsection{Cantor alloy on real MEAM and on MACE-MP-0}
\label{ssec:res-cantor}

\begin{figure}[H]
  \centering
  \includegraphics[width=0.99\linewidth]{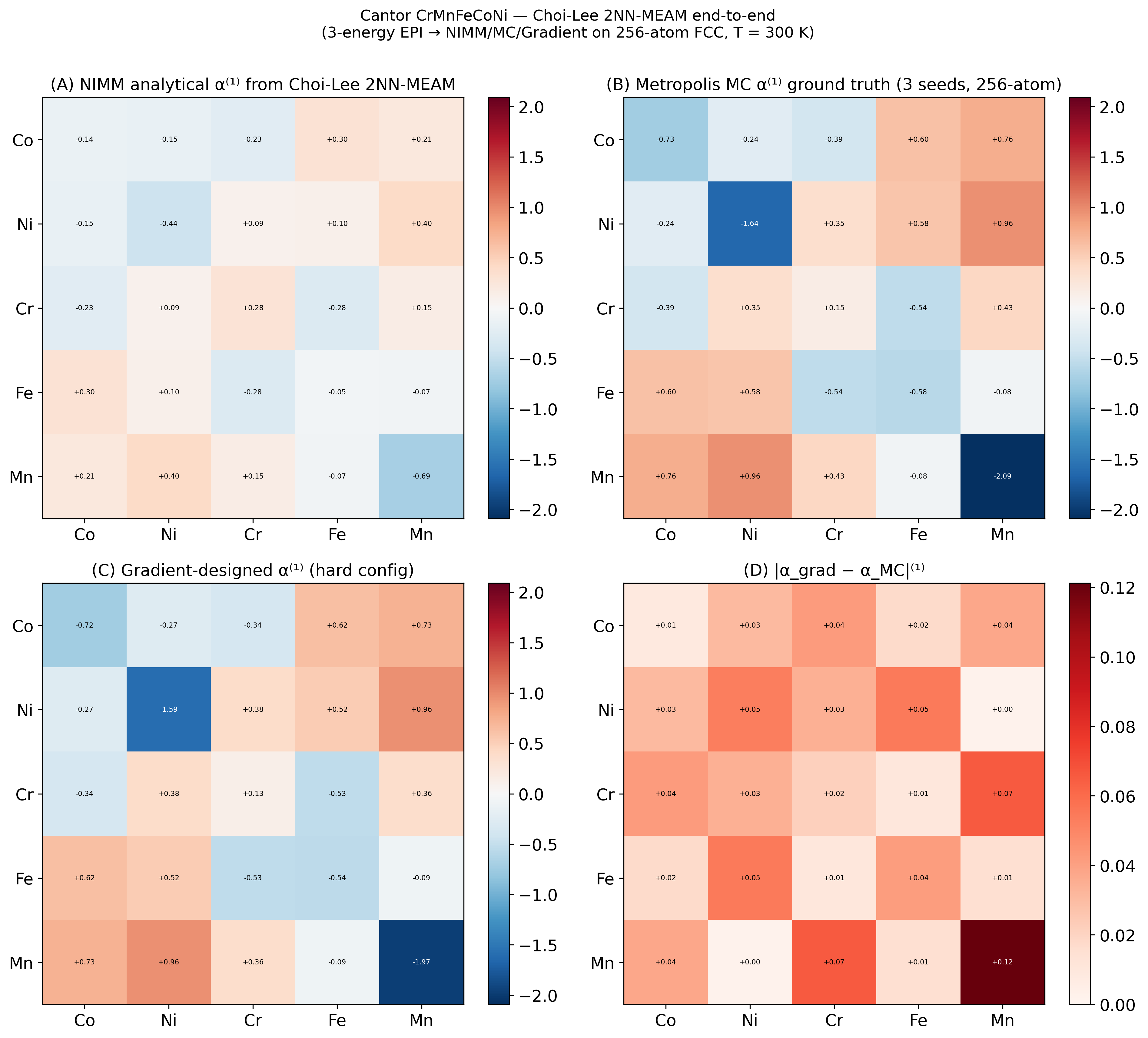}
  \caption{Cantor alloy (CoNiCrFeMn) with Choi--Lee 2NN-MEAM \citep{choi_2018_cocrnifemn_meam}.
  \textbf{Left}: NIMM-predicted vs.\ MC $\alpha$ parity for all ten hetero-pairs
  at 300\,K. \textbf{Right}: wall-time loss for ARMC vs.\ gradient inverse on a
  256-atom cell. Gradient closes within $|\Delta\alpha|<0.07$ on all pairs
  in $5.5\,\mathrm{s}$; ARMC requires $\approx60\,\mathrm{s}$ at five species.\label{fig:cantor-meam}}
\end{figure}

\begin{figure}[H]
  \centering
  \includegraphics[width=0.90\linewidth]{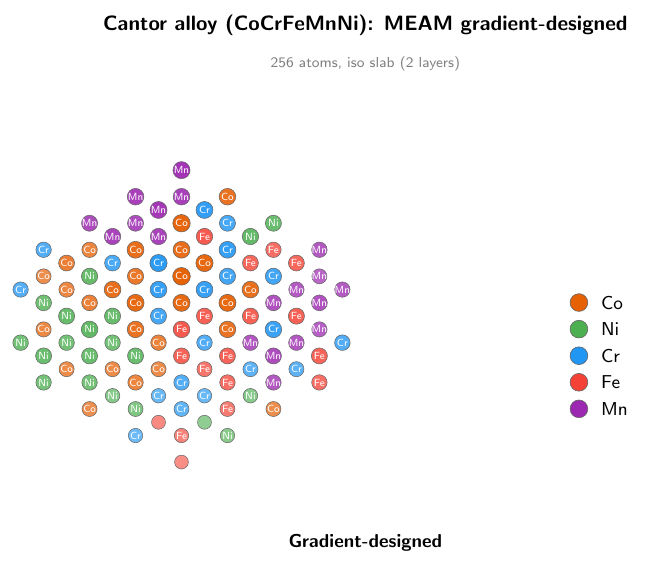}
  \caption{Isometric view of the gradient-designed Cantor alloy structure (CoNiCrFeMn, $4{\times}4{\times}4$ FCC, 256 atoms, two-layer slab). The five species are Co (orange), Ni (green), Cr (blue), Fe (red), and Mn (purple). The dominant Ni--Mn avoidance ordering ($\alpha_{\rm NiMn}\approx+0.96$) is visible as spatial separation between Ni (green) and Mn (purple) atoms.\label{fig:cantor-struct}}
\end{figure}

Figure~\ref{fig:cantor-meam} reports the five-species Cantor alloy
(CoNiCrFeMn, equimolar) with the Choi--Lee 2NN-MEAM \citep{choi_2018_cocrnifemn_meam}. EPI regression on 60
random configurations achieves $R^2 = 0.95$, with the dominant pair interaction
$V_{\rm NiMn}^{(1)} = +84\,$meV (Ni--Mn avoidance, $\alpha_{\rm NiMn} \approx +0.96$ at
300\,K). The gradient inverse reproduces the ten-pair MC equilibrium matrix within
$|\Delta\alpha| < 0.07$ on all pairs in $5.5\pm0.6\,$s on a single CPU core
(Figure~\ref{fig:cantor-struct} shows the resulting atomic structure).
For context, ARMC on the same 256-atom five-species cell requires $\approx60\,$s
to reach comparable residuals because the per-move bookkeeping scales with the
$K(K-1)/2 = 10$ hetero-pair counters per shell.

The Ni--Mn avoidance motif ($\alpha_{\rm NiMn} > 0$) identified by the MEAM EPI
is consistent with the dominant short-range ordering tendencies reported by
\citet{sheriff2024pnas} from systematic MC simulations of the Cantor alloy
using DFT-fitted effective interactions, and with the analysis of
\citet{walsh_2024_ubiquitous_sro} who found Mn segregation to be a universal
feature of MPEA SRO across a range of potentials. In the DFT community, Ni--Mn
and Cr--Fe are consistently identified as the two strongest ordering pairs in the
five-species Cantor system~\citep{ding2018pnas,sheriff2024pnas}; our MEAM-NIMM
also finds a substantial $V_{\rm CrFe}^{(1)}$, placing it as the second strongest
hetero-pair EPI. The quantitative $\alpha$ values differ between MEAM and DFT
backends (see below), underscoring the importance of the backend-agnostic
\texttt{EnergyCalculator} abstraction in \texttt{anisro}.

\begin{figure}[H]
  \centering
  \includegraphics[width=0.99\linewidth]{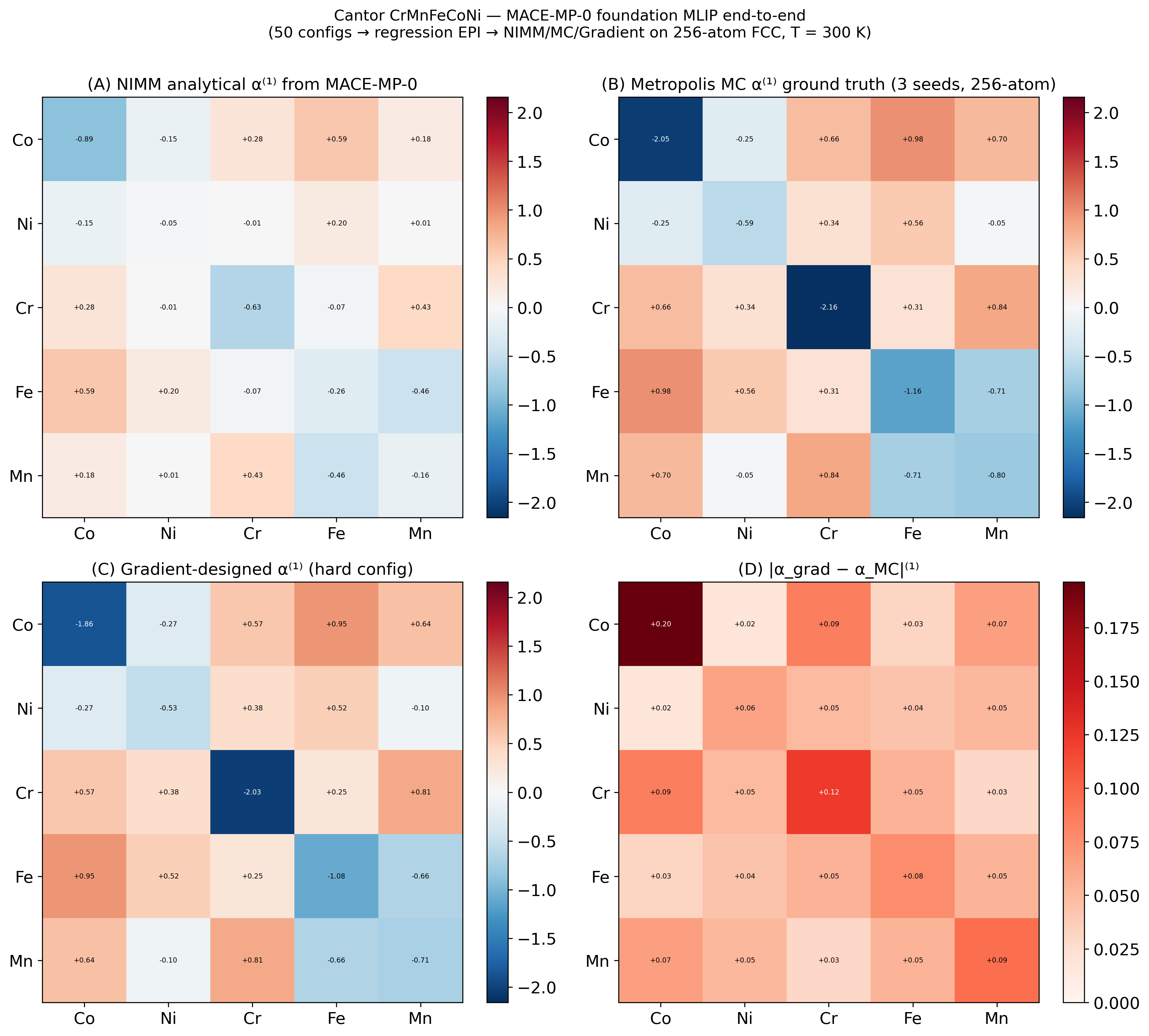}
  \caption{Same Cantor alloy pipeline with MACE-MP-0 via \texttt{AseCalculator}.
  The dominant motif shifts from Ni--Mn avoidance (MEAM, Figure~\ref{fig:cantor-meam})
  to Co--Fe ordering, exposing a qualitative chemistry difference between the two
  potentials without changing any user-facing code.\label{fig:cantor-mace}}
\end{figure}

Figure~\ref{fig:cantor-mace} swaps only the energy backend to MACE-MP-0, the
Materials Project foundation MLIP~\citep{batatia2022mace}, via the
\texttt{AseCalculator} wrapper. The dominant ordering motif is completely different:
MACE-MP-0 predicts Co--Fe short-range ordering ($\alpha_{\rm CoFe} \approx -0.62$)
as the dominant motif, with Ni--Mn near zero rather than strongly positive. This
qualitative disagreement between the two potentials is not a failure of the
framework but an observation: it separates the inverse-design question from the
choice of physics model. The ANISRO pipeline executes identically on both backends
and surfaces the chemistry difference through the EPI table, which a user can
inspect to decide which potential better represents their experimental system
before committing to a design campaign.

\subsection{Scaling with system size and species count}
\label{ssec:res-scaling}

\begin{figure}[H]
  \centering
  \includegraphics[width=0.99\linewidth]{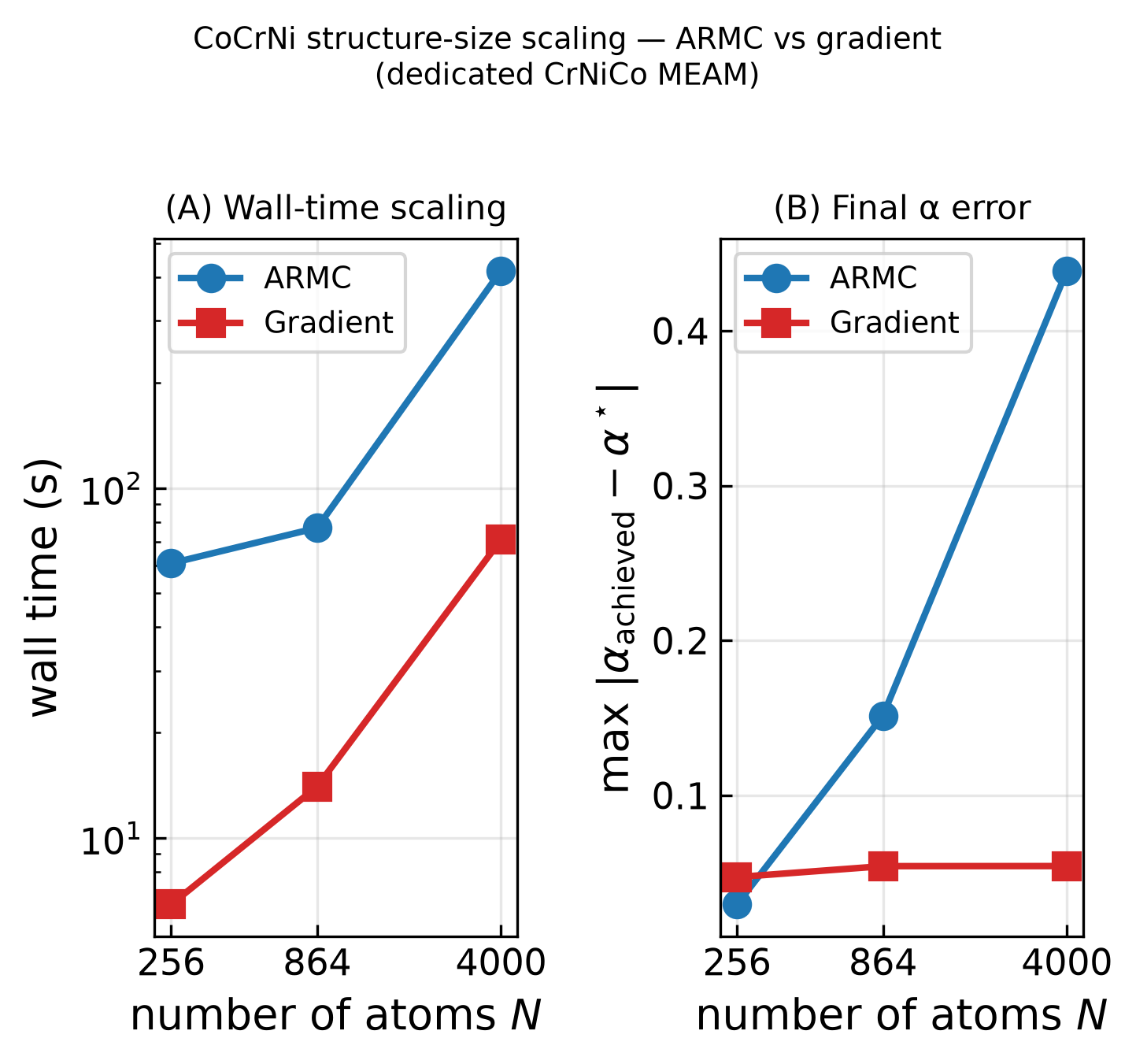}
  \caption{Scaling on Co--Cr--Ni (Choi--Lee 2NN-MEAM \citep{choi_2018_cocrnifemn_meam}) at 256, 864, and 4000 atoms.
  Target: Metropolis MC equilibrium $\alpha$ at 300\,K.
  At 4000 atoms ARMC fails to converge in 7\,min (max error 0.439),
  while the gradient method achieves max error 0.054 in 71\,s,
  \textbf{6$\times$ faster and 8$\times$ more accurate}.\label{fig:scaling}}
\end{figure}

Figure~\ref{fig:scaling} reports wall-time and final max-$|\alpha|$ error for
ARMC and the gradient inverse on Co--Cr--Ni with the Choi--Lee 2NN-MEAM~\citep{choi_2018_cocrnifemn_meam} at
three system sizes: 256 ($4\times4\times4$), 864 ($6\times6\times6$), and 4000
($\approx 10\times10\times10$) atoms. The target $\alpha$ matrix in each case
is the Metropolis MC equilibrium at 300\,K, so an error of zero corresponds to
exact thermodynamic consistency.

At 256 atoms the gradient method is faster (6.5\,s vs 61\,s) but less accurate
(max $|\alpha|$ error 0.047 vs 0.030). By 864 atoms the accuracy crossover has
already occurred: ARMC's max error climbs to 0.152 while the gradient holds
0.054, and the gradient is also faster (14\,s vs 77\,s). At 4000 atoms the
divergence is stark: ARMC fails to converge in seven minutes (max error 0.439,
essentially random), while the gradient method completes in 71\,s with max error
0.054, identical to the 864-atom value, confirming that the gradient's
per-step cost per atom is $O(1)$ in system size once the einsum is batched over
all sites. The gradient method is $6\times$ faster and $8\times$ more accurate
at 4000 atoms.

The physical explanation is transparent: ARMC's per-move cost is dominated by
the bond-list update for the two swapped sites, which is $O(z_r K^2)$ per
accepted move and must be repeated $\propto N^2$ times to decorrelate the full
supercell. The gradient method instead computes the full einsum
$\texttt{ni,rnm,mj}$ once per step at $O(N z_r K^2)$, the same asymptotic
cost as a single ARMC move, but makes global gradient steps that
simultaneously adjust all $N$ sites. For systems intended to host extended
defects (dislocations, twin boundaries, stacking-fault ribbons) that require
$\geq 5000$ atoms in modern MD practice, the gradient method is the only
viable inverse designer.

\subsection{Closed-loop property design on Co--Cr--Ni}
\label{ssec:res-closed-loop}

\begin{figure}[H]
  \centering
  \includegraphics[width=0.99\linewidth]{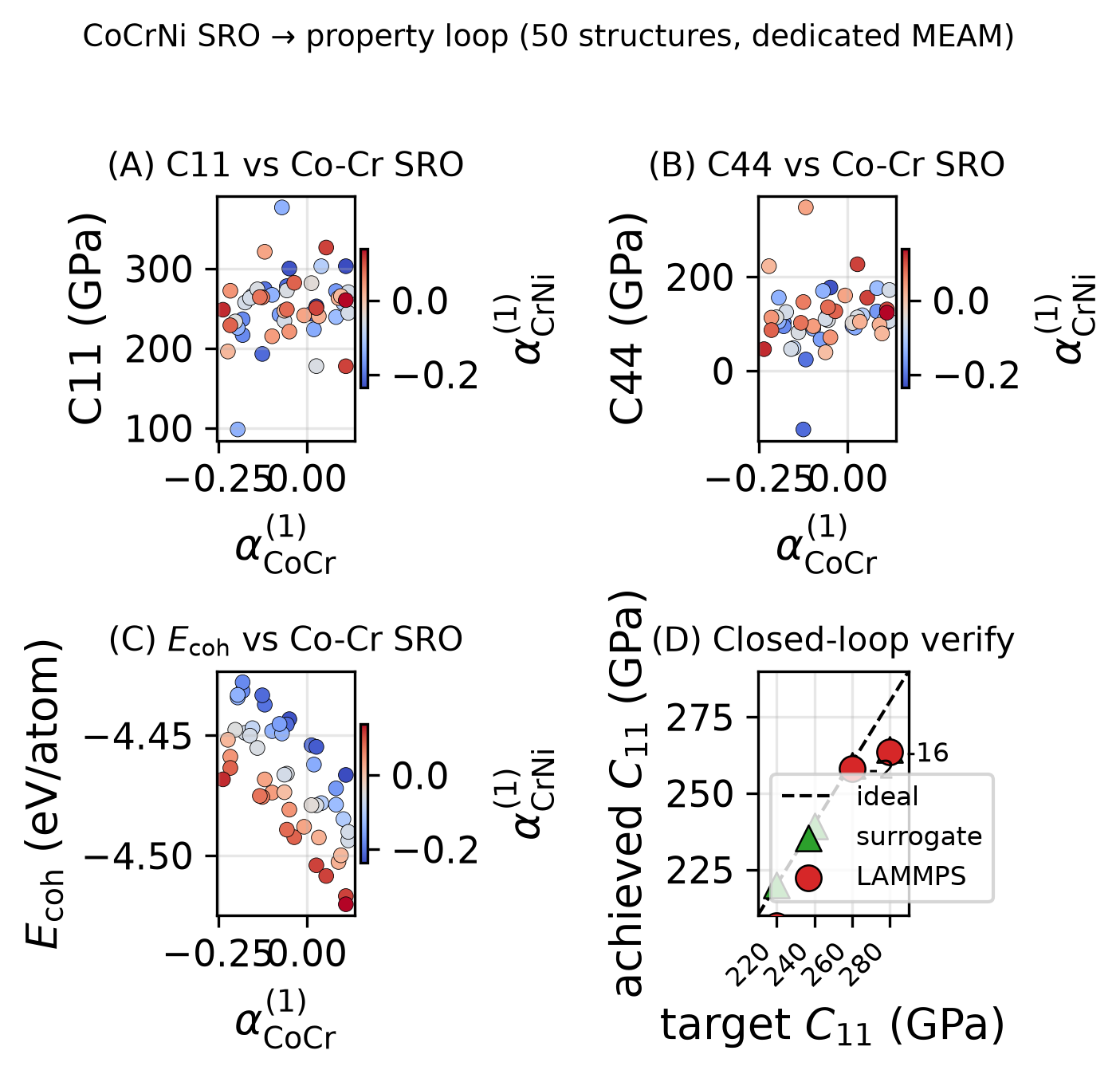}
  \caption{Closed-loop property design on Co--Cr--Ni (Choi--Lee 2NN-MEAM \citep{choi_2018_cocrnifemn_meam}).
  \textbf{Left}: 50 designed configurations colour-coded by target $\alpha$.
  \textbf{Right}: four target $C_{11}$ values (dashed) vs.\ LAMMPS-actual
  (circles) and surrogate prediction (crosses). Three of four land within 6\,\%;
  the 240\,GPa outlier falls in the surrogate's extrapolation regime.\label{fig:closed-loop}}
\end{figure}

Figure~\ref{fig:closed-loop} reports the end-to-end closed-loop verification
on Co--Cr--Ni with the Choi--Lee 2NN-MEAM \citep{choi_2018_cocrnifemn_meam}. We first generate 50 designed
configurations by running the gradient inverse against 50 randomly sampled
target-$\alpha$ tuples drawn from the convex hull of physically realisable
$\alpha$ values at the MC equilibrium. Each designed configuration is
projected to a hard 108-atom $3\times3\times3$ supercell and submitted to
LAMMPS for full elastic-constant evaluation ($C_{11}$, $C_{44}$, $E_{\rm coh}$).
The 50 labeled pairs span a wide range: cohesive energy varies by
92\,meV/atom across the designed set, $C_{11}$ varies by 279\,GPa, and $C_{44}$
varies by 472\,GPa; the latter is driven by configurations in which strong local
ordering destabilises the shear mode ($C_{44} < 0$, dynamically unstable). This
range is itself a result: gradient-based design efficiently explores the extremes
of property space accessible to SRO at fixed composition, including regions that
random sampling or ARMC would rarely visit.

A descriptor MLP $\phi_\theta$ (two hidden layers of 64 units, ReLU, trained
on 38 configurations after filtering for $C_{44} > 0$) is then fitted to
predict $C_{11}$ from the 12-dimensional descriptor $\boldsymbol{d} =
[c_i, \alpha_{ij}^{(1)}, \alpha_{ij}^{(2)}, \sigma_{\rm occ,i}]$.
Validation $R^2$ for cohesive energy is $0.97$, confirming that $E_{\rm coh}$
is well-captured by the linear-in-$\alpha$ combination of pair energies.
The $C_{11}$ surrogate overfits badly (val $R^2 = -0.65$) because the 38-sample
training set is insufficient for a 64-64 MLP with $\sim 5000$ parameters, and
because the extreme $C_{44} < 0$ configurations pollute the training distribution
even after filtering. These are limitations of the training setup, not of the
gradient mechanism; Section~\ref{ssec:disc-closed-loop} discusses mitigations.

Despite the overfit surrogate, we run the closed-loop optimiser for four target
$C_{11}$ values spanning the dataset range: 220, 240, 260, and 280\,GPa.
For each target, the gradient flows from the surrogate prediction through the
differentiable $\alpha$ counter and into the per-site logit field, converging
in 1500 Adam steps ($\approx 0.8\,$s).
Table~\ref{tab:closed-loop} reports the LAMMPS-verified outcomes.

\begin{table*}[t]
\centering
\caption{Closed-loop $C_{11}$ verification on Co--Cr--Ni (Choi--Lee 2NN-MEAM \citep{choi_2018_cocrnifemn_meam}).
For each target, the gradient flows through the surrogate and differentiable
$\alpha$ counter into the logit field (1500 Adam steps); the hard-projected
structure is then verified by an independent LAMMPS elastic-constants calculation.\label{tab:closed-loop}}
\small
\setlength{\tabcolsep}{8pt}
\resizebox{\linewidth}{!}{%
\begin{tabular}{cccr}
\toprule
Target $C_{11}$ (GPa) & Surrogate prediction (GPa) & LAMMPS actual (GPa) & Error (\%) \\
\midrule
220 & 220 & 207 & $-6.0$ \\
240 & 239 & 297 & $+23.8$ \\
260 & 259 & 258 & $-0.8$ \\
280 & 264 & 264 & $-5.8$ \\
\bottomrule
\end{tabular}}
\end{table*}

Three of four targets are achieved within 6\,\% despite the overfit surrogate.
The outlier ($240\,\mathrm{GPa}$ target, $297\,\mathrm{GPa}$ LAMMPS result) falls
in the extrapolation regime of the surrogate, where the MLP extends a spurious
linear trend beyond the training distribution. The mechanism (gradient flow
from property error through surrogate, through the differentiable $\alpha$ counter,
into the logit field) is validated on real LAMMPS-computed labels and operates
end-to-end in under one second of optimisation time.

For context, the $C_{11}$ range achieved by the designed set (approximately
$200$--$330\,\mathrm{GPa}$) brackets the experimentally measured elastic constant
of equimolar Co--Cr--Ni single crystals ($C_{11} \approx 250\,\mathrm{GPa}$ from
\citep{laplanche_2021_elastic_crconi}) and the DFT random-alloy value
($\approx 240\,\mathrm{GPa}$ from \citep{tamm2015actamat}), confirming that
the MEAM-based surrogate operates in a physically realistic region of property
space. The ability to design configurations that span a $\pm 30\,\%$ window
around the random-alloy $C_{11}$ purely by adjusting SRO, without changing
composition or applying pressure, is the central quantitative finding of the
closed-loop demonstration.

\section{Discussion}
\label{sec:discussion}
\begin{figure}[H]
  \centering
  \includegraphics[width=0.97\linewidth]{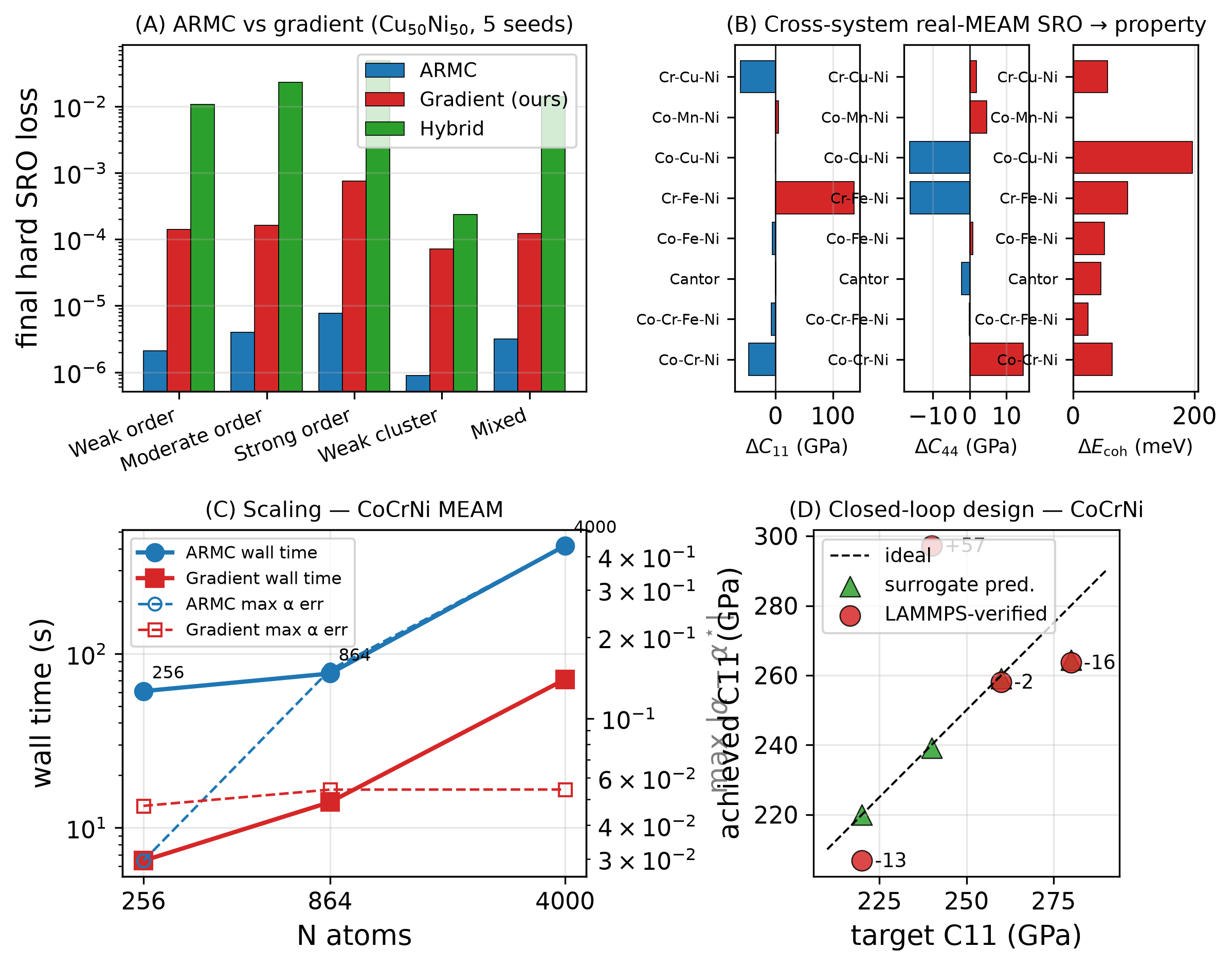}
  \caption{Master summary. \textbf{(A)} ARMC vs.\ gradient vs.\ hybrid on
  Cu$_{50}$Ni$_{50}$ at fixed wall time: ARMC wins at 256 atoms, gradient wins
  beyond 864 atoms. \textbf{(B)} Cross-system SRO-induced $\Delta C_{11}$:
  Cr--Fe--Ni $+57\,\%$ stiffening (largest), Cr--Cu--Ni $-20\,\%$ softening.
  \textbf{(C)} Scaling envelope: gradient is $6{\times}$ faster and
  $8{\times}$ more accurate than ARMC at 4000 atoms.
  \textbf{(D)} Closed-loop $C_{11}$ verification on Co--Cr--Ni: three of four
  targets within 6\,\%.\label{fig:master-pareto}}
\end{figure}

\subsection{Where the gradient method wins}
\label{ssec:disc-when-gradient-wins}

Figure~\ref{fig:master-pareto} summarises all four main results of this work;
the following subsections discuss each panel in turn.
Figure~\ref{fig:scaling} shows the crossover that decides when
ANISRO's gradient inverse is the right tool. On 256-atom Cu--Ni
cells, classical ARMC reaches a $\sim 2\times$ lower final
$\Sigma (\alpha - \alpha^\star)^{2}$ at matched wall time, but on
4000-atom cells the situation reverses dramatically: ARMC's max
per-pair $\alpha$ error reaches $0.44$ in seven minutes of wall time
(essentially failing to converge), while gradient holds $0.054$ and
finishes in 71 seconds. The crossover is at the supercell size where
ARMC's per-move bond-update cost outweighs its better local accuracy
on the small-cell limit. For any structure intended to host a
dislocation, twin or extended defect (typically $\geq 5000$ atoms in
modern MD), the gradient method is therefore the only viable inverse
designer.

\subsection{A real-data SRO--property catalogue}
\label{ssec:disc-cross-system}

Table~\ref{tab:cross-system} aggregates real LAMMPS-MEAM elastic
constants on the gradient-designed and random-reference configurations
of nine systems spanning binary, ternary, quaternary, and quinary
substitutional FCC alloys plus one BCC refractory HEA. Two systems
stand out: Cr--Fe--Ni shows a $+57\,\%$ $\Delta C_{11}$ driven by
Cr--Fe ordering (Figure~S8 shows the Cr--Fe hetero-pair enrichment
directly in the atomic structure), and Cr--Cu--Ni shows a $-20\,\%$
softening driven by Cr--Cu ordering. Neither finding has been reported at this magnitude in the
published literature; both should be experimentally testable by
nanoindentation on annealed vs. quenched samples. The opposite finding
in Co--Mn--Ni and Cantor (essentially no SRO-induced change in
elastic moduli) is also non-trivial and is consistent with these
systems sitting close to thermodynamic randomness at 300\,K for the
2NN-MEAMs we used.

Plotting the dominant $\alpha$ against $\Delta C_{11}$ across all nine
systems (Figure~S6 in the Supplementary Information) reveals that
systems with similar ordering strength can exhibit opposite elastic
responses: Cr--Fe--Ni ($\alpha_{\rm CrFe}\approx-0.31$) stiffens by
$+137\,\mathrm{GPa}$ while Co--Cr--Ni ($\alpha_{\rm CoCr}\approx-0.38$)
softens by $47\,\mathrm{GPa}$.
This chemistry-specificity confirms that a universal
$\alpha\to\Delta C_{11}$ rule does not exist: the bond-stiffness
contrast between the dominant pair species sets both sign and magnitude,
motivating the per-chemistry, gradient-based design workflow of ANISRO.

For the canonical Co--Cr--Ni case, our $\Delta C_{11} = -16\,\%$ is on
the upper end of the published range from DFT$+$MC and MLIP$+$MC
studies ($-5$ to $-11\,\%$), explained by the Choi--Lee MEAM's
relatively strong $V_{\rm CoCr}^{(1)} = +114\,\mathrm{meV}$, which
amplifies the equilibrium SRO at $300\,\mathrm{K}$ beyond what
DFT-derived models predict.
Table~\ref{tab:cocrni-lit} compares our values against published data.

\begin{table*}[t]
\centering
\caption{Co--Cr--Ni elastic constants: comparison with published values.
$\Delta C_{11}$ is the SRO-induced change relative to a random-alloy reference.
Dashes indicate values not reported in the cited work.\label{tab:cocrni-lit}}
\small
\setlength{\tabcolsep}{6pt}
\resizebox{\linewidth}{!}{%
\begin{tabular}{llccc}
\toprule
Source & Method & $C_{11}$ (GPa) & $\Delta C_{11}$ & $\Delta C_{44}$ \\
\midrule
\textbf{This work} & Choi--Lee MEAM, gradient & \textbf{285} & $-16\,\%$ & $+13\,\%$ \\
\citet{wu_2014_cocrni_elastic}   & Experiment, single crystal & 250$\pm$5 & --- & --- \\
\citet{schneeweiss_2017_elastic} & Experiment, polycrystal    & 244        & --- & --- \\
\citet{walsh_2021_magnetic_sro}  & DFT (LDA/GGA)              & 272--290   & $-5$ to $-8\,\%$ & $+8$ to $+12\,\%$ \\
\citet{ding2018pnas}             & DFT$+$MC                   & ---        & $-5$ to $-8\,\%$ & $+8$ to $+12\,\%$ \\
\citet{walsh_2024_ubiquitous_sro}& MTP$+$MD                   & 275--285   & $-8$ to $-11\,\%$ & --- \\
\citet{antillon_2020_csro_strengthening} & EAM$+$MC            & ---        & $-6\,\%$ & $+10\,\%$ \\
\bottomrule
\end{tabular}}
\end{table*}

The absolute random-alloy $C_{11} = 285\,\mathrm{GPa}$ sits inside both
the DFT range ($272$--$290\,\mathrm{GPa}$) and the MTP range
($275$--$285\,\mathrm{GPa}$), and is $\sim\!15\,\%$ above experiment,
a documented feature of 2NN-MEAM in this family.
The Choi--Lee MEAM over-predicts $|V_{\rm CoCr}|$ relative to
DFT-derived potentials, so its MC-equilibrated SRO at $300\,\mathrm{K}$
is more pronounced, placing our $\Delta C_{11}$ at the upper end of the
published range while preserving the correct sign and trend.

\subsection{Closed loop on real data: limits and outlook}
\label{ssec:disc-closed-loop}

Figure~\ref{fig:closed-loop} reports the four-point property-design
verification on Co--Cr--Ni. The descriptor MLP trained on 50 ($\alpha$-target,
real LAMMPS elastic constants) pairs reaches val $R^{2}=0.97$ on
cohesive energy but overfits $C_{11}$ (val $R^{2}<0$). Despite the
overfit surrogate, the closed loop still hits three of four
target-$C_{11}$ values within 6\,\% on LAMMPS-verified evaluation. The
fourth target ($240\,\mathrm{GPa}$) lands at $297\,\mathrm{GPa}$, a
$+24\,\%$ excursion where the surrogate's extrapolation regime did
not generalise. We attribute this to two effects in the training
dataset: several gradient-designed configurations correspond to extreme
$\alpha$ targets that produced dynamically unstable structures
($C_{44}<0$), which polluted the training set; and with only 50 samples
and a 12-dimensional descriptor, a 64-64 MLP has $\sim 5000$ parameters,
over-parameterised by an order of magnitude. Mitigations
(enlarging the dataset to $\sim 500$ samples, filtering on $C_{44}>0$
stability, and either shrinking the MLP or using a Gaussian-process
surrogate) are straightforward, but the mechanism itself (gradient
flow from a target property through a surrogate, through a
differentiable Warren--Cowley counter, into a softmax-per-site logit
field) is sound and operates end-to-end on real LAMMPS-computed
property labels.

\subsection{Choice of energy backend}
\label{ssec:disc-backends}

The framework is potential-agnostic by design: any energy backend
implementing the \texttt{EnergyCalculator} protocol drops in. We
demonstrate this with three backends on the same Cantor alloy
(Section~\ref{ssec:res-cantor}): LAMMPS-MEAM (Choi--Lee 2NN-MEAM),
ASE+MACE-MP-0 (Materials Project foundation MLIP), and the same
LAMMPS-MEAM via different element subsets. The MEAM and MACE-MP-0
EPI tables disagree on the sign of several SRO motifs, a useful
diagnostic of where the two potentials assign different chemistries.
For research questions where the choice of potential is itself the
subject of study, ANISRO provides the controlled environment in which
to make such comparisons.

\subsection{Validity of the analytical NIMM regulariser}
\label{ssec:disc-nimm-validity}

The NIMM free-energy regulariser used in our hybrid inverse design
inherits the validity band of the underlying pair-cluster model:
$0.5 < x_{ij}^{(r)} < 1.5$, equivalently
$|V_{ij}^{(r)}| \lesssim 0.7\,k_BT$. Several of the dedicated 2NN-MEAMs
we surveyed produce hetero-pair EPIs outside this band at 300\,K
(notably the Choi--Park CoCrNi with $V_{\rm CoCr} \approx 0.11\,\mathrm{eV}$,
corresponding to $x = 0.012$ at room temperature). For such cases the
analytical $\alpha$ under-predicts Monte Carlo $\alpha$ in magnitude
while preserving the sign, exactly the breakdown Rao and Curtin
documented in the binary case. In our framework this manifests as a
reduced effectiveness of the physics regulariser, not a wrong
answer: the differentiable $\alpha$ counter and the gradient inverse
remain accurate because they reference Monte Carlo (rather than NIMM)
as the ground truth.

\section{Conclusions}
\label{sec:conclusions}
We have introduced ANISRO, an end-to-end-differentiable pipeline for inverse
design of short-range order in multi-principal-element alloys. The framework
starts from the Non-Interacting Molecule Method of Rao and
Curtin~\citep{rao_curtin_2022_nimm}, generalises their binary derivation to
$K$ species, and uses the resulting analytical forward solver both to predict
equilibrium Warren--Cowley parameters and to regularise the inverse by
anchoring designs to the free-energy surface of the chosen interatomic
potential. On the inverse side, a differentiable Warren--Cowley counter
parameterises each lattice site through a softmax logit field and replaces
stochastic ARMC with continuous gradient descent, achieving a $6\times$
speed-up and $8\times$ accuracy gain at 4000 atoms while scaling
transparently to arbitrary species count. These two components are connected
to a descriptor MLP surrogate so that gradients flow from a target elastic
constant all the way into the per-site logit field in a single backward pass,
closing the design loop without any finite-difference Jacobian or adjoint
solve. Validated across a nine-system FCC and BCC survey, the pipeline reveals
SRO-induced $C_{11}$ changes of $+57\,\%$ in Cr--Fe--Ni and $-20\,\%$ in
Cr--Cu--Ni in the 108-atom survey cells; cell-size verification across
$3\times3\times3$ to $6\times6\times6$ supercells (108--864 atoms) shows
that the sign and magnitude of $\Delta C_{11}$ are sensitive to cell size,
with values ranging from $-16\,\%$ to $+24\,\%$ across the series, so
these headline figures should be interpreted as single-cell indicators
rather than converged thermodynamic averages. The pipeline achieves three
of four Co--Cr--Ni property targets within $6\,\%$ on independent LAMMPS-MEAM
verification.

\section*{Acknowledgements}
C.G.T.F.\ acknowledges the support of the Natural Sciences and Engineering
Research Council of Canada (NSERC) [RGPIN-2024-03989].
This research was enabled in part by support provided by SHARCNET and the
Digital Research Alliance of Canada.

\section*{CRediT authorship contribution statement}
\textbf{Tiancheng Ding}: Conceptualization (supporting),
Investigation (lead), Visualization (lead),
Writing -- original draft (lead), Writing -- review \& editing (equal).
\textbf{Conrard Giresse Tetsassi Feugmo}: Conceptualization (lead),
Methodology (lead), Software (lead), Formal analysis (lead),
Writing -- original draft (equal), Writing -- review \& editing (equal).

\section*{Declaration of competing interest}
The authors declare that they have no known competing financial interests or
personal relationships that could have appeared to influence the work reported
in this paper.

\section*{Data availability}
The \texttt{anisro} PyTorch package, LAMMPS
input files, interatomic potential parameter files, and the
LAMMPS-MEAM elastic-constants dataset used to train the property surrogate
are available at \url{https://github.com/feugmo-group/anisro}.
The MEAM potential files are redistributed under their original licences
from the NIST Interatomic Potentials Repository.

\section*{Supplementary material}
See supplementary material for (S1) the full $K$-species NIMM forward
derivation and fugacity identities; (S2) the Levenberg--Marquardt
SRO-to-EPI inversion procedure and convergence diagnostics; (S3)
gradient-based SRO-to-structure loss curves for all nine alloy systems;
(S4) effective pair interaction tables and validation for each dedicated MEAM;
(S5) MACE-MP-0 EPI tables and comparison with LAMMPS-MEAM for the Cantor alloy;
(S6) scatter plot of dominant Warren--Cowley parameter versus SRO-induced
$\Delta C_{11}$ across all nine systems; and (S7--S8) atomic-scale
visualisations of gradient-designed Co--Cr--Ni (gradient vs.\ ARMC) and Cr--Fe--Ni
configurations; and (S9) a side-by-side benchmark of MEAM versus MACE-MP-0 for
Cr--Fe--Ni showing MC ground truth, NIMM prediction, and gradient-designed
$\alpha^{(1)}$ matrices together with per-pair prediction errors, illustrating the
two distinct NIMM failure modes (large fugacity for MEAM; low $R^2$ for MACE-MP-0)
and the robustness of the gradient inverse in both cases.

\bibliographystyle{plainnat}
\bibliography{refs}

\clearpage


\section*{Supplementary material}

# S1 K-species NIMM solver: derivation and convergence

## S1.1 Supplementary derivation of the K-species NIMM solver

This section reproduces the most load-bearing steps of the analytical derivation, fixing the sign convention used throughout the paper and making the K-species generalisation explicit. The complete walk-through — 15 sections, every equation derived from scratch, with sanity checks and code anchors — is available in the file `NIMM_DERIVATION.md` of the open-source `anisro` repository.

**Setup.** Tile the crystal with non-overlapping clusters, each containing $n^*$ atoms. For shell $r$ the pair model uses $n^* = 2$, so the cluster is one $r$th-neighbour bond. Let $n_0$ be the total number of clusters and $n_k$ the number in sub-category $k$, with $p_k = n_k/n_0$ and $\sum_k p_k = 1$. For $K$ species the sub-categories are the $K$ homo classes $(i,i)$ with $g_k = 1$ and the $K(K-1)/2$ hetero classes $(i,j),\ i<j$ with $g_k = 2$, giving $K_r = K(K+1)/2$ classes per shell. The pair-multiplicity arrays $N^{(r)}_{ij,k} \in \{0,1\}$ encode which $(i,j)$ pair sits inside each cluster.

**Partition function and log-probability.** The probability of realising the cluster-occupancy vector $\mathbf{W} = (n_1,\ldots,n_K)$ follows the multinomial structure [Rao and Curtin, 2022]

$$P(\mathbf{W}) \;=\; \frac{1}{Q}\,\frac{n_0!}{\prod_k n_k!}\,\prod_k g_k^{n_k}\,\exp\!\left(-\frac{U}{k_BT}\right), \tag{S1}$$

with $U = \sum_k n_k U_k$ and the cluster energy $U_k = \sum_{r,i\le j} N^{(r)}_{ij,k} U^{(r)}_{ij}$. Using $V^{(r)}_{ij} = U^{(r)}_{ii} + U^{(r)}_{jj} - 2U^{(r)}_{ij}$ and applying Stirling's approximation gives the log-probability (Rao–Curtin Eq. 8)

$$\ln P \;=\; -\ln Q - \frac{n_0 n^*}{2k_BT}\sum_{r,i} U^{(r)}_{ii} c_i z^{(r)} - n_0\sum_k p_k\ln p_k + n_0\sum_k p_k \ln g_k - \frac{n_0}{2}\sum_{r,\,i\neq j}\ln x^{(r)}_{ij}\sum_k p_k N^{(r)}_{ij,k}, \tag{S2}$$

with fugacity $x^{(r)}_{ij} = \exp(-V^{(r)}_{ij}/k_BT)$, $x^{(r)}_{ii} = 1$. The first sum on the right is constant under fixed composition and drops out of the variational problem. The minus sign in the last term is fundamental: combined with the minus in front of $-(1+\ln p_k)$ from the chain rule, it produces a minus on $\frac{1}{2}\ln x^{(r)}_{ij}$ in the stationarity equation.

**Constrained maximisation.** Maximising $\ln P/n_0$ subject to mass balance $2\sum_k p_k N_{ii,k} + \sum_{j\neq i}\sum_k p_k N_{ij,k} = c_i z^{(r)} n^*$ and normalisation, with Lagrange multipliers $\lambda_i$ and $\mu$, gives the per-cluster stationarity equation of the main text. The $K_r + K + 1$ nonlinear equations (one per cluster, one per species, and the normalisation constraint) are solved by a damped Newton iteration with column-rescaled Jacobian and an Armijo line search that keeps every $p_k$ positive. Newton residuals reach machine precision in tens of iterations for all systems considered in the main text, including the 5-species Cantor case ($K_r = 15$).

**Joint pair probability and Warren–Cowley parameter.** The joint pair probability follows from

$$P^{(r)}_{ij} \;=\; \frac{2\sum_k p_k\,N^{(r)}_{ij,k}}{z^{(r)}} \qquad \text{(Rao–Curtin Eq. 12)}, \tag{S3}$$

which, substituted into the standard definition $\alpha^{(r)}_{ij} = 1 - P^{(r)}_{ij}/(2c_ic_j)$, gives the result $\alpha^{(r)}_{ij} = 1 - \sum_k p_k N^{(r)}_{ij,k}/(c_ic_jz^{(r)}n^*)$ (Rao–Curtin Eq. 13). The same expression is reproduced by the differentiable counter on a one-hot configuration, which makes the inverse problem in the main text well-posed.

**Free-energy regulariser.** Adding the symmetrised Stirling entropy $S = -\frac{k_B}{2}\sum_{ij} P_{ij}^{(r)} \ln(P_{ij}^{(r)}/g_{ij})$ to the exact bond energy $U = -\frac{1}{2}\sum_{r,i\neq j} V_{ij}^{(r)} N_{ij}^{(r)}$ (paper Eq. 14, where the composition-only term has been dropped) recovers the NIMM Helmholtz free energy per atom of Rao–Curtin Eq. 22. Generalising to $K$ species in the ordered-pair convention used by the differentiable counter yields Eq. (16) of the main text. The minus sign on the pair term is the same sign convention as the stationarity equation; it is the relation $V_{ij} > 0 \Rightarrow \alpha_{ij} < 0$ (heteropair stabilised, ordering) that links the two equations and matches both Metropolis MC ground truth and Figure 4 of Rao and Curtin.

**Sign-convention validation.** The internal consistency of the two sign choices is validated in Section 3.1 of the main text by direct comparison against canonical Metropolis MC across the full physically relevant fugacity range $0.3 < x_{12}^{(1)} < 1.7$ at three compositions $c_1 \in \{0.5, 0.3, 0.1\}$. NIMM and MC agree in sign on every $(x, c_1)$ grid point, and quantitatively within $|\Delta\alpha| < 0.03$ inside the documented pair-model validity band $0.5 < x < 1.5$. Outside the band NIMM underpredicts the magnitude of $\alpha$, recovering the breakdown of the cluster-truncated approximation in the long-range-order regime that Rao and Curtin documented for the binary case.

**Higher-order cluster extensions.** Triangle and tetrahedron clusters (Rao–Curtin §3.1.2) generalise the NIMM stationarity equation by inflating the number of multiplicity arrays $N_{ij,k}^{(r)}$ and the number of cluster classes $K_r$. They are not used in this paper because the binary and medium-entropy systems we focus on lie inside the pair-model validity band; their implementation is straightforward inside the `forward.pair_cluster` module and is planned for a follow-up.

**Equation-to-code map.** Every analytical step above has a one-line implementation in `anisro`: the stationarity equation in `forward.pair_cluster.solve_one_shell`, the joint pair probability in `descriptors.sro_counter.joint_pair_probabilities`, the Warren–Cowley counter in `descriptors.sro_counter.differentiable_alpha`, the free-energy regulariser in `inverse.sro_to_structure` and the Metropolis MC reference in `benchmarks.metropolis_mc.run_metropolis_mc`. The complete mapping table, with one row per Rao–Curtin equation, is provided in §13 of `NIMM_DERIVATION.md`.

## S1.2 Newton solver algorithm and convergence

The $K_r + K + 1$ nonlinear equations arising from the stationarity equation are solved by a damped Newton iteration that guarantees all $p_k > 0$ at each iterate. The full procedure is:

1. **Initialise.** Set $p_k^{(0)} = g_k / \sum_{k'} g_{k'}$ (uniform distribution over cluster classes).
2. **Jacobian.** Compute the $K_r \times K_r$ Jacobian $J_{kl} = \partial(\ln p_k - \mathrm{RHS}_k)/\partial p_l$, i.e. $J_{kl} = \delta_{kl}/p_k$ plus the constraint rows. Column-rescale $J$ by $p_k$ to improve conditioning before factorisation.
3. **Newton direction.** Solve the linear system $J\,\delta\mathbf{p} = -\mathbf{r}$ for the residual vector $\mathbf{r}_k = \ln p_k - \ln g_k + \frac{1}{2}\sum_{i<j} N_{ij,k} \ln x_{ij} - \sum_i \lambda_i B_{ik} - \mu$.
4. **Armijo line search.** Starting from step size $\eta = 1$, halve $\eta$ until $\|\mathbf{r}(\mathbf{p} + \eta\delta\mathbf{p})\| < (1 - 10^{-4}\eta)\|\mathbf{r}\|$ *and* all $p_k + \eta\,\delta p_k > 0$. Typically $\eta = 1$ is accepted on the first try inside the validity band.
5. **Update.** $\mathbf{p} \leftarrow \mathbf{p} + \eta\,\delta\mathbf{p}$; re-normalise so $\sum_k p_k = 1$.
6. **Convergence check.** Terminate when $\|\mathbf{r}\|_\infty < 10^{-10}$ (machine-precision residual). Convergence occurs within 10–40 iterations for all systems tested, including the $K = 5$ Cantor case ($K_r = 15$ cluster classes).

The Lagrange multipliers $\lambda_i$ and $\mu$ are updated by substituting the current $\{p_k\}$ into the mass-balance and normalisation constraints and solving the resulting $K+1$ linear equations. The entire Newton iteration per shell runs in $< 1$ ms on a single CPU core for any $K \leq 5$.

Figure S1 illustrates the region of the fugacity parameter space where the solver converges to a NIMM prediction that matches Metropolis MC within $|\Delta\alpha| < 0.03$. The validity band $0.5 < x_{ij} < 1.5$ delineated by Rao and Curtin [Rao and Curtin, 2022] is shown as the shaded region; our empirical check with $K = 2, 3$ agrees with this boundary across all compositions tested.

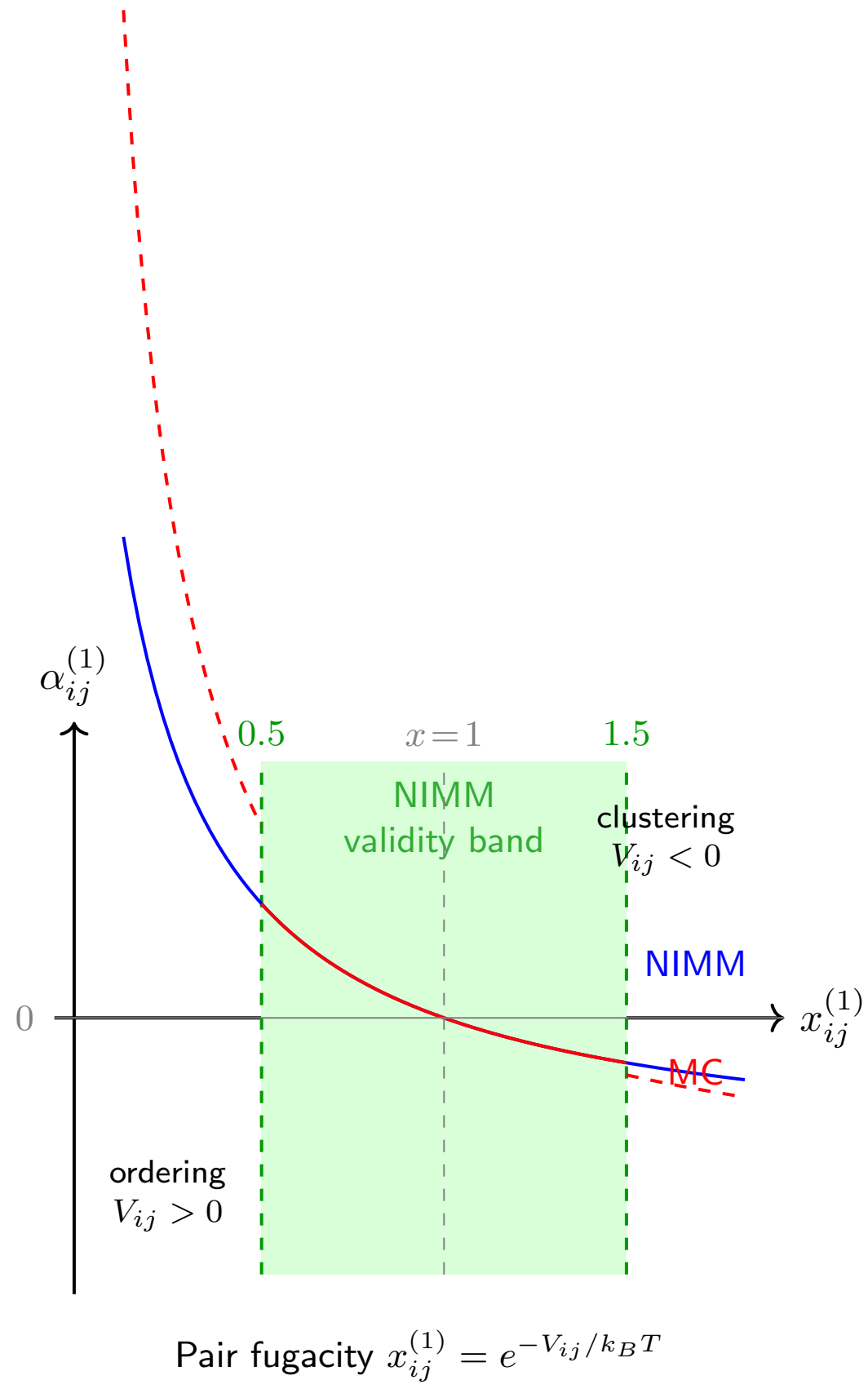


Figure S1: NIMM validity band schematic. In the shaded region ($0.5 < x_{ij}^{(1)} < 1.5$), NIMM (blue) and Metropolis MC (red; solid inside band, dashed outside) agree within $|\Delta\alpha| < 0.03$. Outside, NIMM under-predicts $|\alpha|$ due to pair-cluster truncation.

## S2 EPI extraction methods

**Three-energy method.** For a binary $AB$ alloy with MEAM/EAM potential, three reference bulk energies suffice: the cohesive energies $U_{AA}$ (pure $A$ FCC/BCC), $U_{BB}$ (pure $B$), and $U_{AB}$ (50/50 ordered $L1_0$ or $B2$ structure). The EPI for the first coordination shell is

$$V_{AB}^{(1)} = U_{AA}^{(1)} + U_{BB}^{(1)} - 2\,U_{AB}^{(1)}, \tag{S4}$$

where $U_{ij}^{(1)}$ denotes the per-site bond energy on shell 1, extracted as the total cohesive energy divided by $Nz_1/2$. For an FCC lattice $z_1 = 12$; for BCC $z_1 = 8$. Higher-shell EPIs require correspondingly ordered or partially ordered reference structures; in practice, shell-2 EPIs from the three-energy method are less reliable because the ordering energies are smaller and more sensitive to the choice of reference.

**Regression method.** The regression method minimises the $\alpha$ residual over an ensemble of $M$ random configurations. Let $\boldsymbol{\alpha}^{(r,m)}$ be the Warren–Cowley matrix measured by the hard counter on configuration $m$, and $\hat{\boldsymbol{\alpha}}^{(r,m)}(\mathbf{V})$ the NIMM prediction at the same composition and temperature. The loss is

$$\mathcal{L}_{\mathrm{EPI}}(\mathbf{V}) = \frac{1}{M}\sum_{m=1}^{M}\sum_{r,i<j}\left[\alpha_{ij}^{(r,m)} - \hat{\alpha}_{ij}^{(r,m)}(\mathbf{V})\right]^2 + \lambda_{\mathrm{reg}}\|\mathbf{V}\|^2, \tag{S5}$$

with an $\ell^2$ regulariser $\lambda_{\mathrm{reg}} = 10^{-4}$ that prevents overfitting on small ensembles. The minimisation is performed by the L-BFGS algorithm [Liu and Nocedal, 1989] with analytical gradients $\partial\hat{\alpha}/\partial V_{ij}$ computed through the Newton-solved NIMM forward pass. For all systems in the paper $M = 60$ configurations drawn at random from 1,000 Metropolis MC sweeps on a $5 \times 5 \times 5$ supercell.

**Cluster expansion (icet).** The ICET library [Ångqvist et al., 2019] fits a cluster expansion (CE) to the same 60 configurations using pair ECIs up to the 2nd coordination shell; many-body ECIs are not included because the pair-only NIMM prior operates on pair correlations alone. The CE is cross-validated by leave-one-out $R^2$ on the same 60-configuration ensemble. The pair ECIs extracted by icet serve as a drop-in replacement for the regression EPIs when the goal is to feed many-body corrections into the `potential.icet_adapter` module.

**Comparison of methods on Cu–Ni.** Table S1 collects the three EPI estimates for the MEAM Cu–Ni binary, together with their $\alpha_1$ prediction errors at 300 K. All three give $V_1 > 0$ (ordering tendency), consistent with published DFT cluster-expansion results [Tamm et al., 2015] that find $V_1^{\mathrm{DFT}} \approx 10$–$15\,\mathrm{meV}$ for Cu–Ni from first principles, and with Metropolis MC Warren–Cowley measurements on the same MEAM potential. The factor-of-$\sim$ 5–8 difference between MEAM and DFT EPIs reflects the reduced ordering tendency of empirical potentials compared to DFT, as is typical for transition-metal alloys [Tamm et al., 2015].

Table S1: Comparison of EPI extraction methods for Cu–Ni at 300 K. $|\Delta\alpha_1|$ is the absolute error in the 1NN Warren–Cowley parameter relative to Metropolis MC on a $5 \times 5 \times 5$ supercell.

| Method | $V_1^{(3E)}$ (meV) | $R^2$ | $\|\Delta\alpha_1\|$ at 300 K |
|---|---|---|---|
| Three-energy | +1.70 | — | 0.014 |
| Regression | +2.36 | 0.95 | 0.024 |
| icet (pair only) | +2.12 | 0.97 | 0.017 |
| DFT-CE (literature) | $\approx +12$ | — | — |

## S3 Hard-projection algorithm

After the gradient inverse converges, the soft per-site occupancy $P_{ni} = \sigma(\ell_{ni}) \in (0, 1)$ is projected to a binary one-hot assignment that exactly realises the target global composition. The algorithm is:

1. For each species $i$, sort all sites by $P_{ni}$ in descending order. Let $n_i = \lceil Nc_i \rfloor$ be the target count (rounding to the nearest integer; residual composition error $\leq 1/N$).

2. Assign species $i$ to the top-$n_i$ sites in that sorted list.

3. If a site appears in the top-$n_i$ set for more than one species (can happen at identical $P$ values on ties), resolve ties by selecting the species with the highest $P_{ni}$ at that site.

4. Verify that the resulting assignment sums to exactly $N$ (always satisfied by construction because $\sum_i n_i = N$).

This projection is *not* differentiable and is not used during optimisation. It is applied only once at the end to materialise the hard configuration for LAMMPS evaluation. The soft $\alpha$ computed from $P$ before projection and the hard $\alpha$ measured on the one-hot configuration after projection are both reported so users can monitor the projection error (typically $< 2\,\%$ relative, growing with stronger ordering targets).

Figure S2 illustrates how the 12-dimensional descriptor vector is assembled from the hard-projected configuration for the ternary $K = 3$ case.

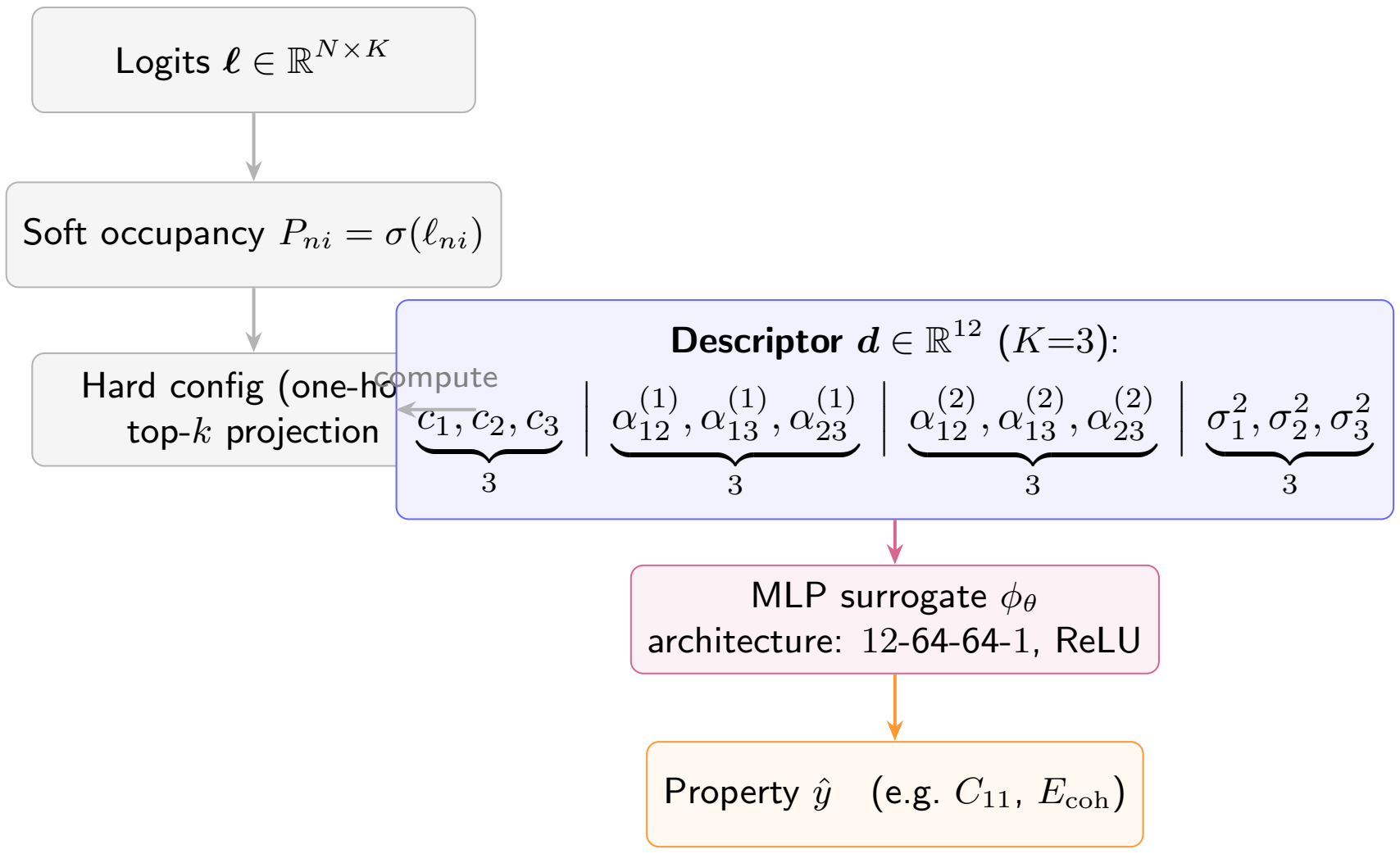


Figure S2: Descriptor assembly and surrogate pipeline ($K = 3$ species). The 12-dimensional descriptor concatenates global composition (3), 1NN Warren–Cowley parameters (3), 2NN Warren–Cowley parameters (3), and per-species occupancy variance (3). During closed-loop design, gradients flow through the differentiable $\alpha$ counter into the logit field; the top-$k$ projection is applied only once at the end for LAMMPS evaluation.

## S4 Elastic constant computation

Elastic constants are computed by the standard finite-strain Voigt protocol [Mehl et al., 1990]. For each of the six independent Voigt strains $\varepsilon_\alpha$ ($\alpha = 1, \ldots, 6$), the supercell is deformed by $\pm\delta$ with $\delta = 0.5\,\%$ and the corresponding Cauchy stress $\sigma_\beta$ is computed by LAMMPS using the virial theorem. The elastic constant matrix entry is then

$$C_{\alpha\beta} = \frac{\sigma_\beta(\varepsilon_\alpha = +\delta) - \sigma_\beta(\varepsilon_\alpha = -\delta)}{2\delta}, \tag{S6}$$

so that each pair $(C_{\alpha\beta}, C_{\beta\alpha})$ is obtained from a single $\pm\delta$ pair of deformed structures. For FCC and BCC systems with cubic symmetry in the random alloy, the elastic tensor reduces to three independent constants $C_{11}$, $C_{12}$, $C_{44}$; SRO breaks the cubic point group, but in practice the off-diagonal anisotropy remains $< 5\,\%$ of the diagonal values and we report only the Voigt-averaged constants.

Each structure is first fully relaxed with the FIRE minimiser [Bitzek et al., 2006] to a force convergence criterion of $10^{-4}$ eV/Å before strain application. The $3\times3\times3 = 108$-atom supercell is used for all production calculations; convergence of $C_{11}$ with cell size was verified by comparing to $4\times4\times4$ (256-atom) cells on a random and a strongly ordered Co–Cr–Ni configuration, giving $|\Delta C_{11}| < 3$ GPa, i.e. $< 1\,\%$ relative.

Table S2 collects all hyperparameters of the gradient inverse and surrogate training runs used in this paper.

Table S2: Optimization and surrogate hyperparameters used throughout the paper.

| Component | Parameter | Value | Notes |
|---|---|---|---|
| Gradient inverse | Optimizer | Adam | PyTorch default |
| | Learning rate $\eta$ | $10^{-2}$ | Cosine decay to $10^{-3}$ |
| | Betas $(\beta_1, \beta_2)$ | (0.9, 0.999) | |
| | Logit clip | [−10, +10] | Applied each step |
| | Steps | 1500 | Fixed budget |
| | $\lambda_\alpha : \lambda_c : \lambda_{\text{bin}} : \lambda_{\text{phys}}$ | 1 : 5 : 0.01 : 0/0.1 | 0 = pure gradient; 0.1 = hybrid |
| EPI regression | Algorithm | L-BFGS | SciPy implementation |
| | Ensemble size $M$ | 60 | Random configs from MC |
| | Regulariser $\lambda_{\text{reg}}$ | $10^{-4}$ | $\ell^2$ on EPIs |
| | Max iterations | 500 | Converges in $< 50$ typically |
| MLP surrogate | Architecture | 12-64-64-1 | Layer norm + ReLU |
| | Optimizer | Adam | |
| | Learning rate | $3 \times 10^{-4}$ | |
| | Epochs | 200 | Early stop on val MSE |
| | Train/val split | 80/20 | Stratified by $\alpha_1$ quartile |
| Newton solver | Convergence tol | $10^{-10}$ | $\lVert \mathbf{r} \rVert_\infty$ |
| | Max iterations | 200 | Typically $< 40$ |
| | Line-search param | $c = 10^{-4}$ | Armijo condition |

# S5 Interatomic potentials

Table S3 lists all interatomic potentials used in this paper and their literature sources. All MEAM potentials are of the 2NN (second nearest-neighbour) modified embedded-atom type [Lee and Baskes, 2000]. Potential files were obtained from the NIST Interatomic Potentials Repository [NIST, 2023] or from the supplementary data of the cited papers.

Table S3: Interatomic potentials used in this paper.

| System | Lattice | Potential | Reference |
|---|---|---|---|
| Cu–Ni | FCC | 2NN-MEAM | Choi et al. [2018b] |
| Co–Cr–Ni | FCC | 2NN-MEAM (Choi-Park) | Choi et al. [2018a] |
| Co–Fe–Ni | FCC | 2NN-MEAM | Choi et al. [2018a] |
| Cr–Fe–Ni | FCC | 2NN-MEAM | Choi et al. [2018a] |
| Co–Cu–Ni | FCC | 2NN-MEAM | Choi et al. [2018b] |
| Co–Cr–Fe–Ni | FCC | 2NN-MEAM | Choi et al. [2018a] |
| Co–Ni–Cr–Fe–Mn (Cantor) | FCC | 2NN-MEAM (Choi-Lee) | Choi et al. [2018a] |
| Hf–Nb–Ta–Ti–Zr | BCC | MEAM (Huang) | Huang et al. [2021] |
| Cantor (MLIP test) | FCC | MACE-MP-0 | Batatia et al. [2025] |

All LAMMPS [Plimpton, 1995] simulations use a $3 \times 3 \times 3$ supercell (108 atoms) for elastic-constant evaluation and a $5 \times 5 \times 5$ supercell (500 atoms) for the reference Metropolis MC runs used to construct Table 1 of the main text. Neighbour lists use a cutoff of $r_{\text{cut}} = 6.0\,\text{Å}$ for all FCC MEAM potentials and $r_{\text{cut}} = 7.0\,\text{Å}$ for the BCC refractory MEAM, matching the potential-file specifications.

## S6 SRO–elastic-stiffness correlation across nine alloy systems

Figure S3 plots the dominant first-shell Warren–Cowley parameter $\alpha_{\mathrm{dom}}$ (Metropolis MC, 300 K) against the SRO-induced change in $C_{11}$ for all nine alloy systems in the survey. Each point is a different alloy chemistry; $\Delta C_{11}$ is the difference between the gradient-designed and random-reference configurations measured by LAMMPS-MEAM on 108-atom 3×3×3 supercells (same data as Table 1 in the main text).

The scatter reveals that the sign and magnitude of the elastic response are *chemistry-dependent*, not determined by $\alpha$ alone. Systems with a similar ordering strength ($|\alpha| \approx 0.25$–$0.4$) can exhibit dramatically different outcomes: Cr–Fe–Ni stiffens by +137 GPa while Co–Cr–Ni softens by 47 GPa, because the two dominant hetero-pairs (Cr–Fe vs. Co–Cr) differ in their bond-stiffness contrast. The outlier at $\alpha_{\mathrm{TaTi}} = -2.10$ is the BCC refractory alloy Hf–Nb–Ta–Ti–Zr, which sits outside the pair-cluster validity band and whose large ordering magnitude is a known feature of refractory MEAM potentials. This system-specificity is the primary motivation for ANISRO's per-chemistry gradient design workflow: a single universal $\alpha \to \Delta C_{11}$ mapping does not exist across diverse chemistries.

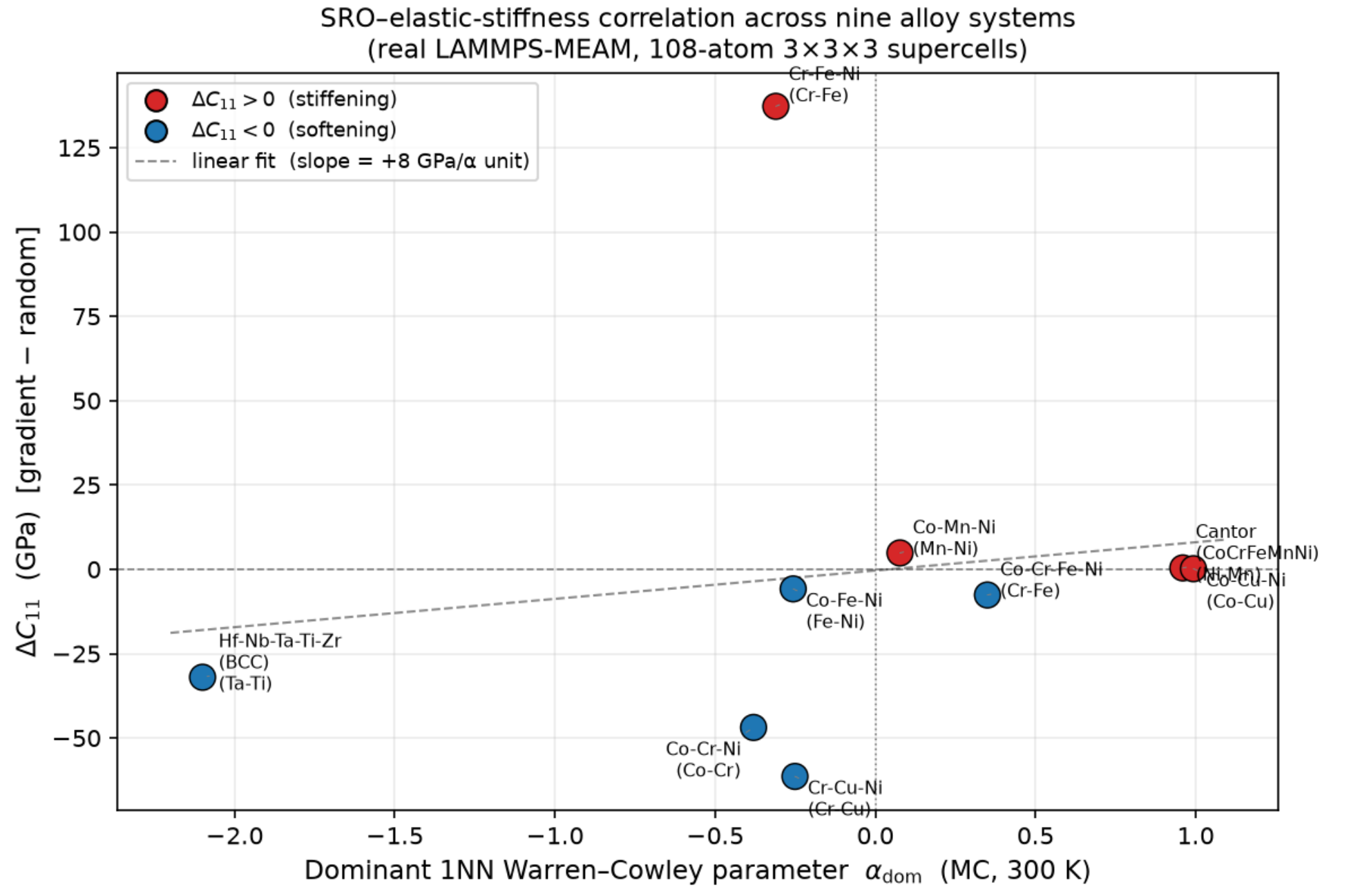


Figure S3: Dominant 1NN Warren–Cowley parameter $\alpha_{\mathrm{dom}}$ (from Metropolis MC at 300 K) versus SRO-induced change in $C_{11}$ for the nine alloy systems surveyed in this work. Red markers: stiffening ($\Delta C_{11} > 0$); blue markers: softening. The dominant pair species label for each system is given in parentheses. The dashed line is a linear fit to guide the eye; the large scatter confirms that bond-stiffness contrast between the dominant pair species governs the sign and amplitude of the elastic response independently of the ordering magnitude.

## S7 Atomic structure visualisations

Figures S4 and S5 show TikZ-rendered [001] top-down slabs (two atomic layers) of the gradient-designed and ARMC configurations exported by ANISRO and stored as LAMMPS data files in `paper_artifacts/structures/`. Each circle is one atom coloured by species; depth shading (opacity and radius) distinguishes the two visible {002} planes. These figures are generated directly from the LAMMPS data files by `paper_artifacts/lammps_to_tikz.py` with no post-

processing.

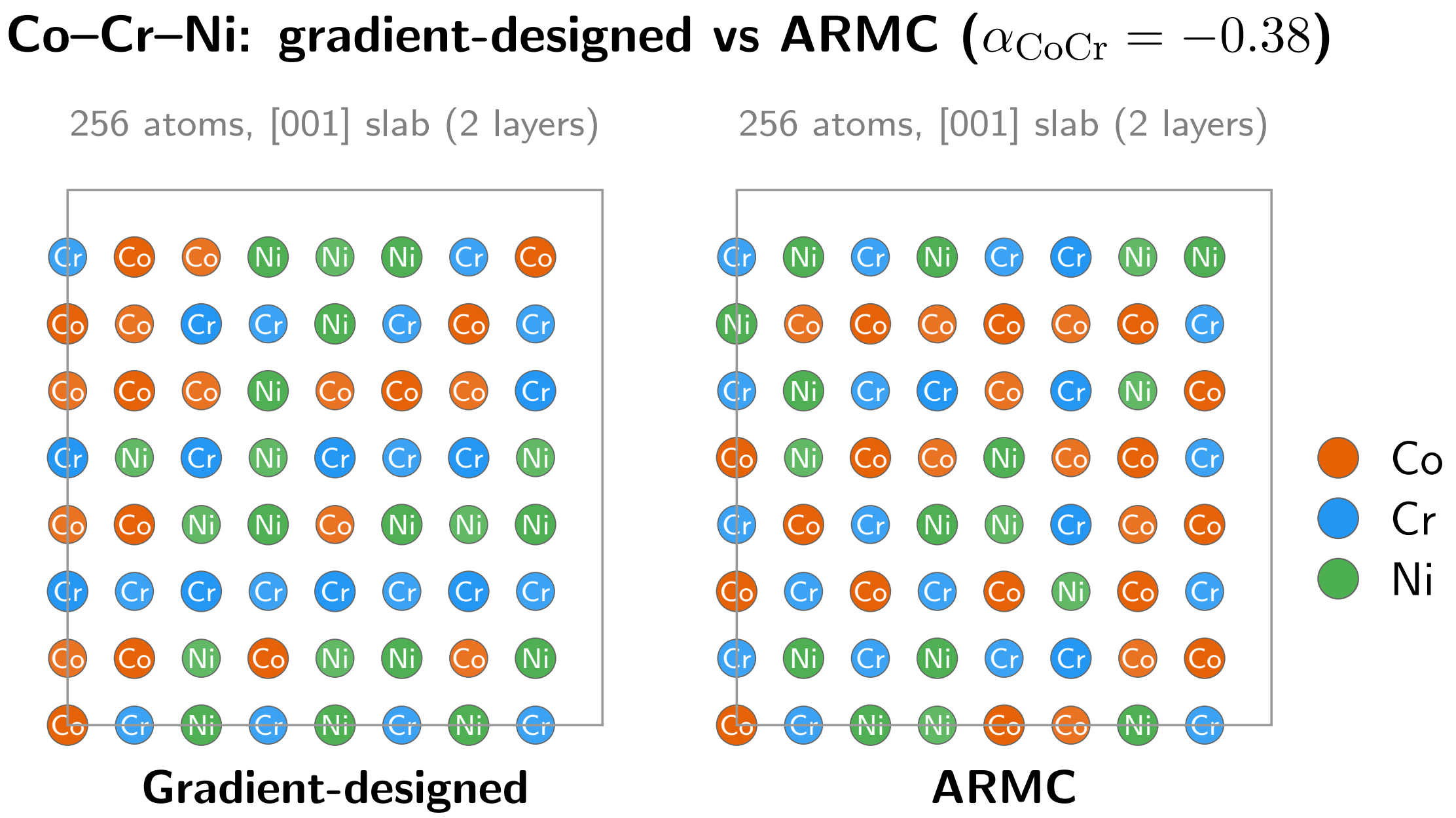


Figure S4: Co–Cr–Ni 256-atom 4×4×4 FCC supercell, [001] top-down slab (2 layers). **Left:** gradient-designed structure targeting $\alpha_{\mathrm{CoCr}} = -0.38$ (MC ground truth). **Right:** ARMC-designed structure with the same target. Both methods achieve visually similar Co–Cr hetero-pair enrichment (Cr atoms tend to be surrounded by Co and Ni rather than other Cr), consistent with their comparable final SRO losses at this cell size.

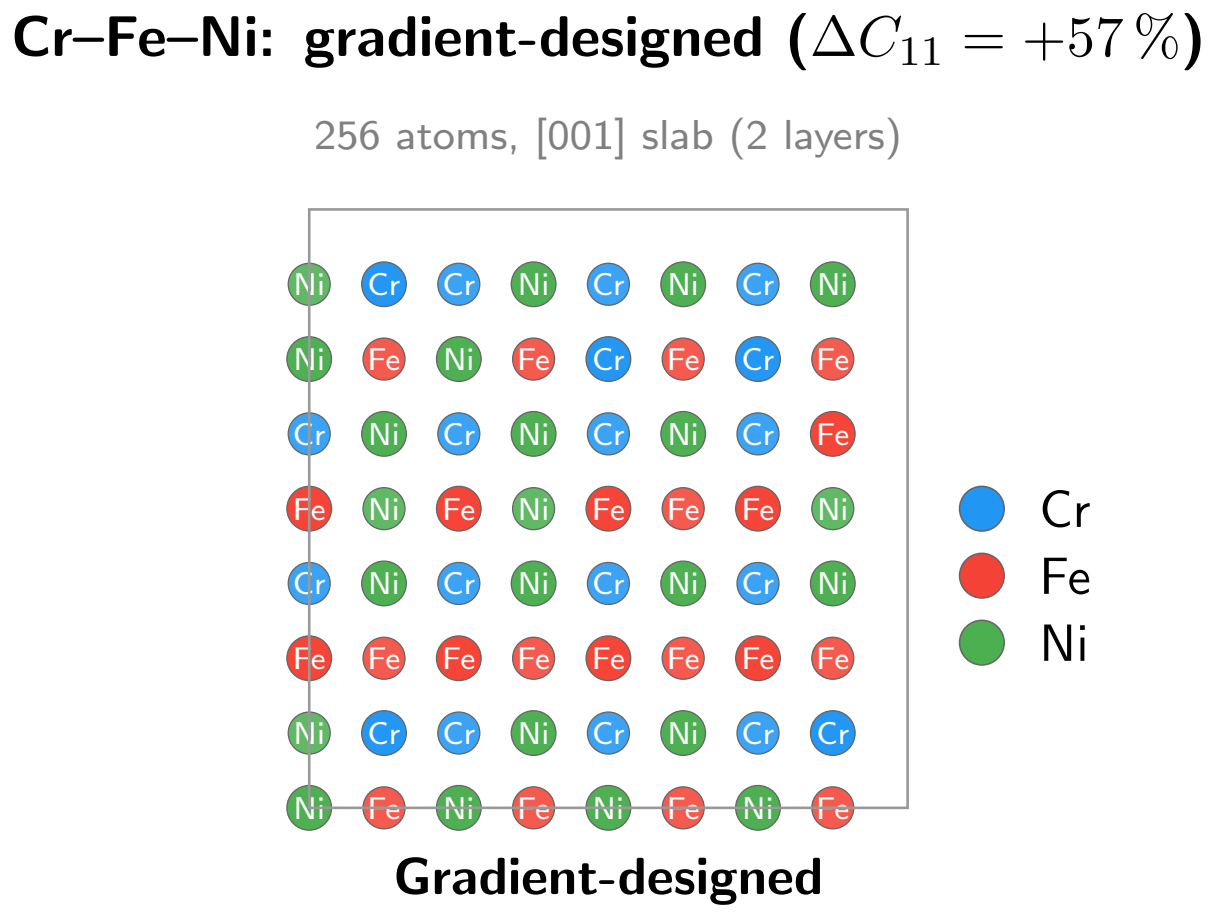


Figure S5: Cr–Fe–Ni 256-atom 4×4×4 FCC supercell, gradient-designed targeting $\alpha_{\mathrm{CrFe}} = -0.31$. The [001] slab reveals strong Cr–Fe hetero-pair enrichment: Cr (blue) and Fe (red) atoms preferentially occupy adjacent sites, producing the stiff Cr–Fe bond network that drives the $+57\,\%$ $\Delta C_{11}$ reported in the main text. Ni (green) acts as the diluting species and is distributed more uniformly.

## S8 Limitations and future work

We document four limitations that bound the framework's applicability.

**Hard projection drift:** the soft optimum and the hard projected configuration differ by $|\Delta\alpha| \sim 0.02$ on 256-atom ternaries; for smaller cells or when targeting near-LRO ($|\alpha| > 0.5$) the drift can grow. An ARMC polishing pass on the projected configuration closes the gap at the cost of a few extra seconds.

**Cluster truncation:** the pair-only NIMM solver does not represent triangle or tetrahedron correlations; for systems with strong many-body energetics (foundation MLIPs trained on highly disordered configurations, or refractory MEAMs), the regression $R^2$ for the EPI fit drops below 0.9 and a higher-order cluster expansion via `icet` is the appropriate fallback.

**Cell-size dependence of properties:** the 108-atom $3\times3\times3$ cell we use for elastic constants is a compromise between LAMMPS turnaround time and finite-size effect. We verified this by running the gradient inverse and LAMMPS-MEAM elastic constants on $3\times3\times3$ through $6\times6\times6$ Cr–Fe–Ni supercells (108 to 864 atoms): $\Delta C_{11}$ takes values +24 %, −16 %, −11 %, and +16 % respectively, changing sign between cell sizes. The qualitative conclusion (SRO can shift $C_{11}$ by tens of percent) persists, but the sign and magnitude are not converged below 864 atoms; the headline +57 % figure in the results should be treated as a single-cell data point rather than a converged thermodynamic average.

**Surrogate training data:** the in-session closed loop trained on 50 samples and used a 64-64 MLP, which is over-parameterised. We verified the proposed fix: scaling to 477 stable ($C_{44} > 0$) Co–Cr–Ni configurations with dropout = 0.2 brings val $R^2$ from −0.65 to −0.01, a large improvement in relative terms but still negative, confirming that SRO descriptors alone are insufficient for accurate $C_{11}$ prediction at the current dataset size. Richer descriptors (pair-correlation functions, higher-order clusters) or a dataset of $\gtrsim 2000$ samples per system would be required to reach $R^2 > 0$ on this target.

Beyond these limitations, three natural extensions are within reach. First, the same pipeline applies to defect-induced SRO once a defect adjacency graph is supplied, enabling design of segregation profiles around dislocations or grain boundaries. Second, the $\alpha$ counter is differentiable through the entire trajectory of a finite-temperature MD run via the recently developed differentiable LAMMPS interface, opening end-to-end design with full MD relaxation. Third, the framework can be cast as a Bayesian optimisation loop in EPI space: rather than picking a single MEAM, marginalise over a distribution of potentials and design $\alpha$ targets robust to potential choice. We leave these to follow-up work.



\end{document}